\documentclass[a4paper,11pt]{article}
\pdfoutput=1 
\usepackage{jcappub}
\usepackage{multirow}
\usepackage[T1]{fontenc}
\usepackage[utf8]{inputenc}
\usepackage{lineno}
\usepackage[normalem]{ulem}

\title{\boldmath Cosmogenic photon and neutrino fluxes in the Auger era}

\author[a]{Rafael {Alves~Batista}}
\author[b]{Rogerio M. de Almeida}
\author[c]{Bruno Lago}
\author[d,e]{Kumiko Kotera}

\affiliation[a]{Universidade de São Paulo, Instituto de Astronomia, Geofísica e Ciências Atmosféricas; Rua do Matão, 1226, 05508-090, São Paulo-SP, Brazil}
\affiliation[b]{Universidade Federal Fluminense, EEIMVR; Volta Redonda-RJ, Brazil}
\affiliation[c]{Centro Federal de Educação Tecnológica Celso Suckow da Fonseca – Campus Nova Friburgo (CEFET/RJ – Campus Nova Friburgo), Nova Friburgo-RJ, Brazil}
\affiliation[d]{Sorbonne Universités, UPMC Univ. Paris 6 et CNRS; UMR 7095, Institut d’Astrophysique de Paris, 98 bis bd Arago, 75014 Paris, France}
\affiliation[e]{Laboratoire AIM-Paris-Saclay, CEA/DSM/IRFU, CNRS, Université Paris Diderot; F-91191 Gif-sur-Yvette, France}

\emailAdd{rafael.ab@usp.br}

\abstract{
The interaction of ultra-high-energy cosmic rays (UHECRs) with pervasive photon fields generates associated cosmogenic fluxes of neutrinos and photons due to photohadronic and photonuclear processes taking place in the intergalactic medium. We perform a fit of the UHECR spectrum and composition measured by the Pierre Auger Observatory for four source emissivity scenarios: power-law redshift dependence with one free parameter, active galactic nuclei, gamma-ray bursts, and star formation history. We show that negative source emissivity evolution is favoured if we treat the source evolution as a free parameter. In all cases, the best fit is obtained for relatively hard spectral indices and low maximal rigidities, for compositions at injection dominated by intermediate nuclei (nitrogen and silicon groups). In light of these results, we calculate the associated fluxes of neutrinos and photons. Finally, we discuss the prospects for the future generation of high-energy neutrino and gamma-ray observatories to constrain the sources of UHECRs.
}

\begin{document}

\maketitle
\flushbottom

\section{Introduction}

Ultra-high-energy cosmic rays (UHECRs) are particles, mostly atomic nuclei, with energies $E \gtrsim 1 \; \text{EeV}$ ($1 \; \text{EeV} \equiv 10^{18} \; \text{eV}$). Neither their origins nor the mechanisms whereby they are accelerated to such high energies have been unveiled.

Data from the Pierre Auger Observatory~\cite{auger2015b} suggest a light composition at $\sim 1 \; \text{EeV}$, which becomes heavier as energy increases, favouring an intermediate-mass composition for the highest-energy events~\cite{auger2016a}. Results from the Telescope Array (TA)~\cite{ta2015a} confirm this trend, but accounting for uncertainties they are also compatible with a predominantly protonic composition.

A variety of mechanisms have been put forth to explain the acceleration of UHECRs by a number of astrophysical sources. Each combination of source and acceleration models lead to a different prediction of the flux of cosmic rays produced. The escape of UHECRs from the magnetised environment of the source is ensured if the Larmor radii of the particles, which depend on the source's magnetisaton, are larger than the typical size of the acceleration site -- the so-called Hillas condition~\cite{hillas1984a}. Possible acceleration sites/events include jets of active galactic nuclei (AGNs), gamma-ray bursts (GRBs), magnetars, young pulsars, tidals disruption events, among others. 

The energy spectrum of the cosmic rays emitted by a source depends on the acceleration mechanism, the strength of the magnetic field near the source, and the density of ambient target fields. Nevertheless, it is often assumed to be a power law with a cutoff that reflects the maximal energy attainable by the accelerator. The accelerators also evolve with redshift, adding an additional layer of complication to the modelling of sources.

It is widely believed that UHECRs have extragalactic origin~\cite{auger2017b}. Thus, they can interact with the intergalactic medium including photon fields such as the cosmic microwave background (CMB) and the extragalactic background light (EBL). Magnetic fields, too, play an important role in UHECR propagation. Because their distribution in the Universe is not well understood, the prospects for ultra-high-energy cosmic-ray astronomy are unclear (see e.g. Refs.~\cite{hackstein2016a,alvesbatista2017a,hackstein2017a}).

The so-called dip model~\cite{berezinsky2005a,berezinsky2006a} postulates a purely protonic composition for the UHECR flux and was once the prevalent paradigm. However, many pure proton models have been disfavoured due to the fact that they overproduce the diffuse gamma-ray background (DGRB)~\cite{berezinsky2011a,gavish2016a,berezinsky2016a,supanitsky2016a,vanvliet2017a} and/or violate neutrino limits~\cite{gelmini2012a,kistler2014a,heinze2016a}. Because protons tend to contribute to the overall cosmogenic fluxes considerably more than other nuclei, even in a mixed-composition scenario it is possible to set limits on the total fraction of protons arriving at Earth based on the associated cosmogenic fluxes~\cite{hooper2011a,vanvliet2017b}. In recent years, much effort has been put into this kind of study~\cite{murase2012b,aloisio2015a,globus2017a}.

Recently the Pierre Auger Collaboration has performed a combined spectrum-composition fit~\cite{auger2017a}. A number of simplified assumptions are made in this work, namely that the sources are uniformly distributed within the comoving volume and that only five nuclear species are emitted (H, He, N, Si, and Fe), being these five species a representative sample that can approximately describe reality. The extragalactic magnetic field is assumed to be null. It was also attempted to account for theoretical uncertainties related to the modelling of propagation and cross check results of two simulations codes, photonuclear cross sections, and models of the EBL.  

The results of the Auger fit are rather surprising, pointing to a ``hard-spectrum problem'', favouring low spectral indices ($\alpha \lesssim 1$), and posing challenges to the current acceleration paradigm. 
The situation can be alleviated by making a distinction between the spectrum of the accelerated particles ($\alpha_\text{acc}$) and the one of the escaping particles ($\alpha_\text{esc}$), since interactions with the gas and photons surrounding a source and the local magnetic fields can drastically change the spectral shape. In the following discussion, unless otherwise stated, we will refer to the \emph{escape} spectra as this is the relevant phenomenological quantity. Low spectral indices are incompatible with most common acceleration models, including Fermi ($\alpha_\text{acc} \approx 2$). Ref.~\cite{unger2015a} suggests that photodisintegration of UHE nuclei in the environment surrounding sources may alter the spectrum, yielding $\alpha_\text{esc} \approx 1$. 
Very hard spectra ($\alpha_\text{esc} \sim 0.3-0.8$) are predicted by some plasma-based models~\cite{heavens1988a}. The recent study~\cite{globus2015a} on UHECR acceleration by internal shocks in gamma-ray bursts (GRBs) has obtained spectral indices as low as $\alpha_\text{esc} \approx 0$. This  substantial spectral hardening has also been observed in the GRB models of Refs.~\cite{baerwald2013a,biehl2018a,zhang2018a}. Acceleration by unipolar inductors taking place in young pulsars produces hard spectra, with $\alpha_\text{esc} \approx 1$~\cite{blasi2000a,fang2012a}. The necessary conditions for acceleration are also realised in magnetic reconnection sites, which can produce $\alpha_\text{esc} \approx 1-1.5$~\cite{kowal2011a,drury2012a,dalpino2014a}. The tidal disruption event (TDE) model of Ref.~\cite{alvesbatista2017b} with ignition of the star, too, may produce hard spectra with indices as low as $\alpha_\text{esc} \sim 1.2$. Refs.~\cite{biehl2017a,zhang2017a,guepin2018a} have also suggested that jetted TDEs may lead to hard spectra. 

Even if hard spectra production were achieved at each source, it is likely that the overall diffuse UHECR spectrum softens through the integration over the population of sources. Indeed, sources with milder characteristics, which will produce lower-energy particles, are more numerous, and the distribution of their parameters will naturally soften the spectrum~\cite{kotera2011a,fang2013a}. A hard spectrum production is thus difficult to justify from a source population point of view.

The suppression of the low-energy part of the UHECR spectrum due to the diffusion of low-energy particles in extragalactic magnetic fields, the magnetic horizon effect, has been proposed as a solution to this problem~\cite{globus2008a,mollerach2013a}. Nevertheless, they do not necessarily play a significant role at $E \gtrsim 10^{18} \; \text{eV}$ if one assumes realistic magnetic field configurations~\cite{lemoine2005a,kotera2008a,alvesbatista2014a,alvesbatista2017a}. Another solution has been put forth by Taylor {\it et al.}~\cite{taylor2015a}, who argue that a negative source evolution can be invoked to reconcile theoretical models with observations. This trend has been confirmed in Ref.~\cite{auger2017a} for uniformly distributed sources, and Ref.~\cite{wittkowski2017a} for sources following the large-scale distribution of matter in the universe. In the present work we extend the fit performed by the Auger Collaboration~\cite{auger2017a,wittkowski2017a} to include a more detailed treatment of the cosmological evolution of the source emissivity distribution. We then use these results to compute fluxes of photons and neutrinos stemming from cosmic-ray interactions. 

Our calculations can be understood as conservative lower limits to the cosmogenic neutrino and photon fluxes. Indeed, as we discuss in section~\ref{sec:conclusions}, by relaxing our symplifying assumptions the flux of cosmogenic particles would necessarily increase. Taking into account the latest Auger results enables us to conservatively narrow down the range of allowed cosmogenic neutrino fluxes considerably, compared to the previous work of Ref.~\cite{kotera2010a}.

This paper is structured as follows: in section~\ref{sec:propagation} we provide the theoretical foundations for the modelling of UHECR propagation, presenting details of the setup of simulations and the code in section~\ref{sec:simulations}; results of the fit are presented in section~\ref{sec:fitResults}; predictions of cosmogenic fluxes in light of the fit results are given in sections~\ref{sec:cosmogenicResultsPh} and~\ref{sec:cosmogenicResultsNu}, for photons and neutrinos, respectively; finally, in section~\ref{sec:conclusions} we conclude and discuss future prospects for detecting cosmogenic neutrinos and photons.

\section{Propagation of UHECRs, gamma rays, and neutrinos}\label{sec:propagation}

UHECRs interact with photons from the CMB and EBL. Cosmogenic photons and neutrinos stem from these interactions.
One of the most important of such process for production of cosmogenic particles is photopion production, which in the case of UHE protons can be written as $p + \gamma_{bg} \rightarrow \Delta^+ \rightarrow n + \pi^+$ and $p + \gamma_{bg} \rightarrow \Delta^+ \rightarrow p + \pi^0$. The neutral pion decays predominantly as $\pi^0 \rightarrow 2 \gamma$, and the channel with the largest branching ratio for the decay of the charged pion is $\pi^+ \rightarrow \mu^+ + \bar{\nu}_\mu \rightarrow e^+ + \nu_e + \bar{\nu}_\mu$. These processes are responsible for the well-known Greisen-Zatsepin-Kuz'min (GZK) cutoff, which sets a limit of $\sim 100 \; \text{Mpc}$ to the maximum distance from which UHE protons can be detected with energies exceeding the GZK threshold ($E_\text{GZK} \approx 4 \times 10^{19} \; \text{eV}$). Beta decay is responsible for generating neutrinos through the decay of the neutron: $n \rightarrow p + e^- + \bar{\nu}_e$.

Likewise, nuclei with atomic number $Z > 1$ can also interact with CMB and EBL photons via photopion production. The interaction rate for this process is approximately a superposition of the corresponding rates for protons and neutrons. 
Bethe-Heitler pair production ($^A _Z X + \gamma \rightarrow ^A _Z X + e^+ + e^-$) generates an electron-positron pair that will contribute to the development of electromagnetic cascades thereby affecting the observed photon flux.

Photodisintegration is a photonuclear process whereby a nucleus is split into smaller components due to interactions with photons. This process is an important channel for photon production at ultra-high energies. In particular, radiative decays of excited states produced in the photodisintegration chain, such as $^A _Z X^* \rightarrow ^A _Z X + q \gamma$, with the asterisk denoting an excited state and $q$ the number of photons produced, can lead to copious amounts of high-energy photons.

Cosmogenic neutrinos produced through the aforementioned processes do not interact and propagate rectilinearly, virtually undisturbed. Photons, on the other hand, interact with the CMB, the EBL, and the universal radio background (URB), producing electromagnetic cascades in the intergalactic medium. The main interaction processes are pair production ($\gamma + \gamma_{bg} \rightarrow e^+ + e^-$) and double pair production ($\gamma + \gamma_{bg} \rightarrow e^+ + e^- + e^+ + e^-$), in the case of photons, and triplet pair production ($e^\pm + \gamma_{bg} \rightarrow e^\pm + e^+ + e^-$) and inverse Compton scattering ($e^\pm + \gamma_{bg} \rightarrow e^\pm + \gamma$) in the case of electrons and positrons. Charged particles can emit synchrotron radiation in the presence of magnetic fields. This contribution is particularly relevant for cascade electrons, but typically small for UHECRs.

\section{Propagation models and simulations}\label{sec:simulations}

To simulate the propagation of UHECRs and cosmogenic photons and neutrinos, we use the CRPropa 3 code~\cite{alvesbatista2016a}. 
We include the most recent developments such as the additional photon production channels introduced in the latest releases~\cite{heiter2018a}. We consider all relevant energy-loss processes and particle interactions, namely: Bethe-Heitler pair production, photopion production, photodisintegration and nuclear decay, as well as adiabatic energy losses due to the expansion of the universe. All interactions relevant for photon and electron propagation are taken into account and implemented in CRPropa following~\cite{lee1998a}.

We assume a distribution of sources whose comoving emissivity evolves with redshift as
\begin{equation}
	\dot{\varepsilon}(z) = \dot{\varepsilon}_0 (1 + z)^m,
	\label{eq:srcDensity}
\end{equation}
wherein $m$ is the source evolution parameter, and $0 \leq z \leq 1$. The emissivity, by definition, accounts for both for the effects of source density and luminosity. Equation~\ref{eq:srcDensity} is a very rough approximation as the evolution of most source candidates tends to change in different redshift intervals. We have also considered three particular cases for source evolution. 

We define the evolution for the star formation rate (SFR) as~\cite{hopkins2006a,heinze2016a}:
\begin{equation}
	\psi_\text{SFR}(z) \propto 
	\begin{cases}
	(1 + z)^{3.44}             & \; \text{if }\, z<0.97 \\
	10^{1.09} (1 + z)^{-0.26}  & \; \text{if }\, 0.97<z<4.48 \\
	10^{6.66} (1+z)^{-7.8}     & \; \text{if }\, z>4.48
	\end{cases}.
	\label{eq:psiSFR}
\end{equation}
For gamma-ray bursts (GRBs) the evolution is defined as~\cite{wanderman2009a}
\begin{equation}
	\psi(z)_\text{GRB} \propto 
	\begin{cases}
	(1 + z)^{2.1}             & \; \text{if }\, z<3 \\
	(1 + z)^{-1.4}			  & \; \text{if }\, z\geq 3 \\
	\end{cases}.
	\label{eq:psiGRB}
\end{equation}

The evolution of AGNs depends on their luminosities, which typically range from $10^{42}$ to $10^{46}$ in units of $h^{-2} \; \text{erg}\,\text{s}^{-1}$, wherein $h$ denotes the normalised Hubble parameter. Lovelace~\cite{lovelace1976a} and Waxman~\cite{waxman2004a} argue that a luminosity $L \gtrsim 10^{44-46} \; \text{erg}\,\text{s}^{-1}$ is required to accelerate particles to ultra-high energies in AGN jets. This excludes the populations of low- and medium-low luminosity AGNs as possible accelerators. Because the number density of AGNs decreases with luminosity, as shown in Ref.~\cite{hasinger2005a}, we choose the evolution of medium-high luminosity AGNs, which reads~\cite{hasinger2005a}
\begin{equation}
	\psi_\text{AGN}(z) \propto 
	\begin{cases}
	(1 + z)^{5.0}   & \; \text{if }\, z<1.7 \\
	\text{constant} & \; \text{if }\, 1.7<z<2.7 \\
	10^{2.7 - z}    & \; \text{if }\, z>2.7
	\end{cases},
	\label{eq:psiAGN}
\end{equation}
which is the same expression used in Ref.~\cite{gelmini2012a}. This applies to AGNs with luminosities $10^{44-45} h^{-2} \; \text{erg}\,\text{s}^{-1}$. For higher luminosities, the evolution is $(1 + z)^{7.1}$ for $z < 1.7$, but the number density of this population is more than an order of magnitude lower than for the ones we have used~\cite{hasinger2005a}. Therefore, we henceforth refer to the population of AGNs whose evolution is given by Eq.~\ref{eq:psiAGN} with luminosity  $\sim 10^{44.5} h^{-2} \; \text{erg}\,\text{s}^{-1}$ simply as AGNs.

Particles are injected by sources with energies between $0.1\;\text{EeV}$ and $1000\;\text{EeV}$. The energy spectrum is modelled as
\begin{equation}
	J(E) = J_0 \sum\limits_{j} f_j  E^{-\alpha}
	\begin{cases}
	1  &  \text{if } Z_j R_\text{max} > E \\[0.6em]
	\exp \left( 1 - \dfrac{E}{Z R_\text{max}} \right) &  \text{if } Z_j R_\text{max} \leq E \\[0.6em]
	\end{cases},
\end{equation}
wherein $j \in \{\text{H},\text{He},\text{N},\text{Si},\text{Fe}\}$ designates the composition of the injected nuclei with atomic number $Z_j$ and abundance $f_j$, $\alpha$ is the spectral index of injection, and $R_\text{max}$ the cutoff rigidity, which reflects the maximal energy attainable by the sources, and $J_0$ is an overall normalisation. 

We adopt the EBL model by Gilmore {\it et al.}~\cite{gilmore2012a}. The URB model used was the one by Protheroe and Biermann~\cite{protheroe1996a}.
Photonuclear cross sections are computed based on TALYS 1.8~\cite{konig2005a}, the default setting of CRPropa. The impact of different EBL models on the propagation of UHECRs is discussed in detail in Ref.~\cite{alvesbatista2015a}, and the uncertainties in cross section are addressed in Refs.~\cite{alvesbatista2015a,boncioli2016a,soriano2018a}. Uncertainties in the propagation, namely the EBL and photodisintegration, are also important and affect the results of the fit, as discussed in Ref.~\cite{alvesbatista2016a}.
It is beyond the scope of this work to include all these theoretical uncertainties; we aim to obtain order-of-magnitude estimates using reasonable conservative assumptions. 

In order to infer the mass composition of the arriving simulated particles we follow the same procedure as Ref.~\cite{auger2017a}. The development of the extensive air shower is modelled with a Gumbel function $g(X_\text{max} | E, A)$~\cite{dedomenico2013a}, using results obtained with CONEX~\cite{pierog2006a}. We have adopted the EPOS-LHC model~\cite{pierog2013a} to estimate the depth of the shower maximum ($X_\text{max}$). We do not consider other models for the sake of simplicity. A more detailed analysis of the effect of different hadronic interaction models on the fit is shown in Ref.~\cite{auger2017a,romerowolf2018a}.

We have simulated events up to $z_\text{max} = 1$. This is because for $E \gtrsim 10^{18.7} \; \text{eV}$ virtually all cosmic rays come from $z \lesssim 1$. Nevertheless, because cosmogenic fluxes depend on the choice of $z_\text{max}$, we have also run additional simulations up to $z_\text{max}=5$, which are discussed in Appendix~\ref{app:zmax}.
\section{Results of the fit}\label{sec:fitResults}

We have performed a fit of the spectrum and composition measured by Auger. The procedure is the same one used in Ref.~\cite{auger2017a}, and is described in Appendix~\ref{app:fitting}. In this section we will use the quantity $\sqrt{D - D_\text{min}}$ as a proxy for the standard deviation.

First, we check our implementation of the fit procedure by comparing our results with Ref.~\cite{auger2017a}. In the limit of no source evolution ($m=0$) we obtain the following best-fit parameters: $\alpha = -1.0$, $\log(R_\text{max} / V) = 18.2$, $f_\text{H} = 0.6726$, $f_\text{He}=0.3135$, $f_\text{N}=0.0133$, and $f_\text{Si}=0.0006$. These numbers are in agreement with the results by the Pierre Auger Collaboration for the corresponding scenario, which are: $\alpha=-1.03^{+0.35}_{-0.30}$, $\log(R_\text{max} / V) = 18.21^{+0.05}_{-0.04}$, $f_\text{H}=0.68$, $f_\text{He}=0.31$, $f_\text{N}=0.01$, and $f_\text{Si}=0.0006$. 

\begin{figure}
  \centering
  \includegraphics[width=.495\columnwidth]{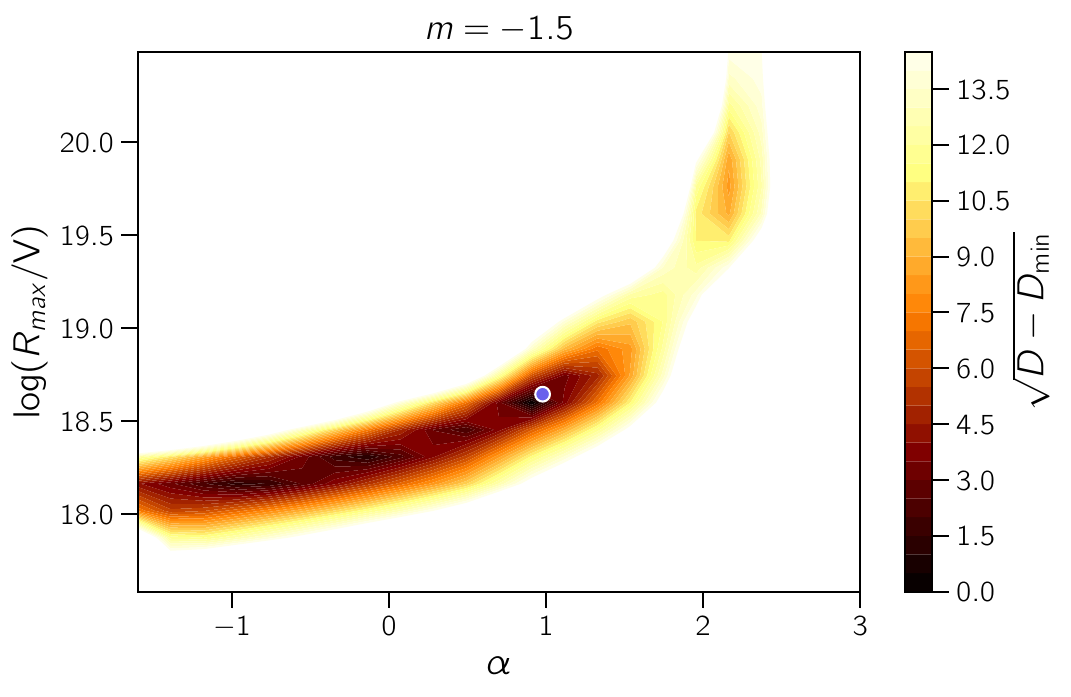}
  \includegraphics[width=.495\columnwidth]{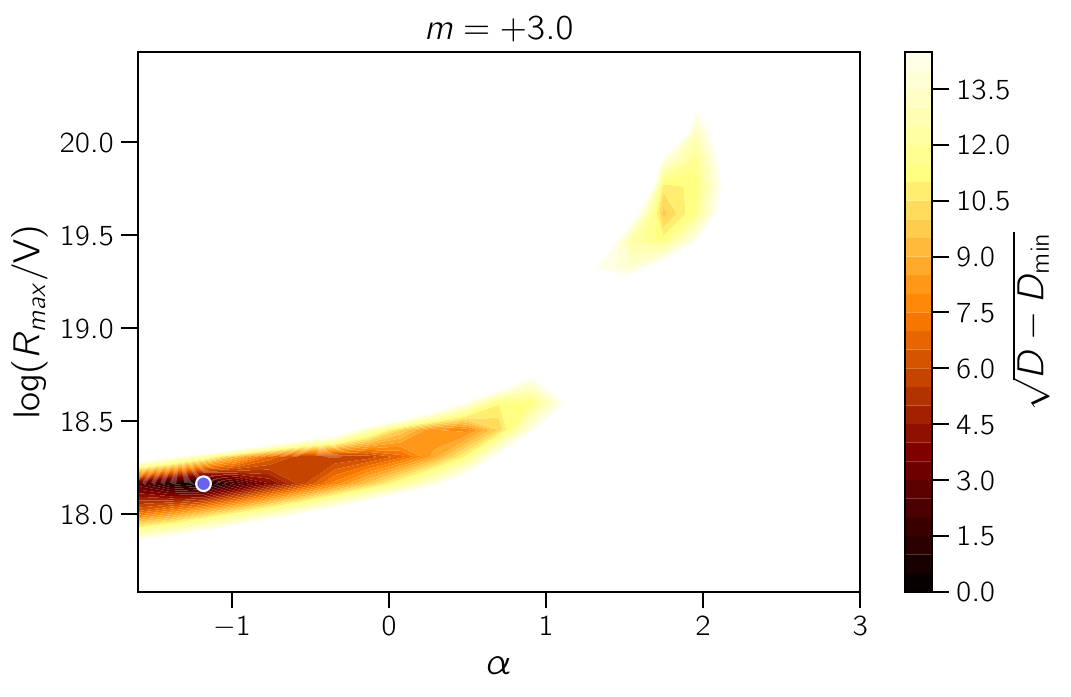}
  \includegraphics[width=.495\columnwidth]{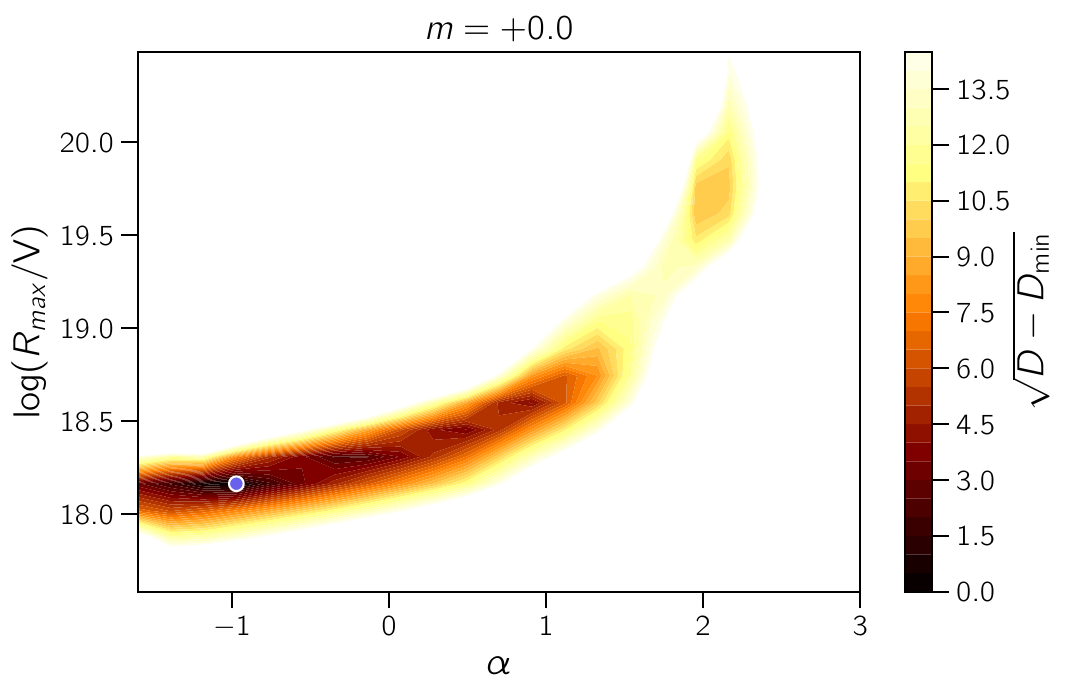}
  \includegraphics[width=.495\columnwidth]{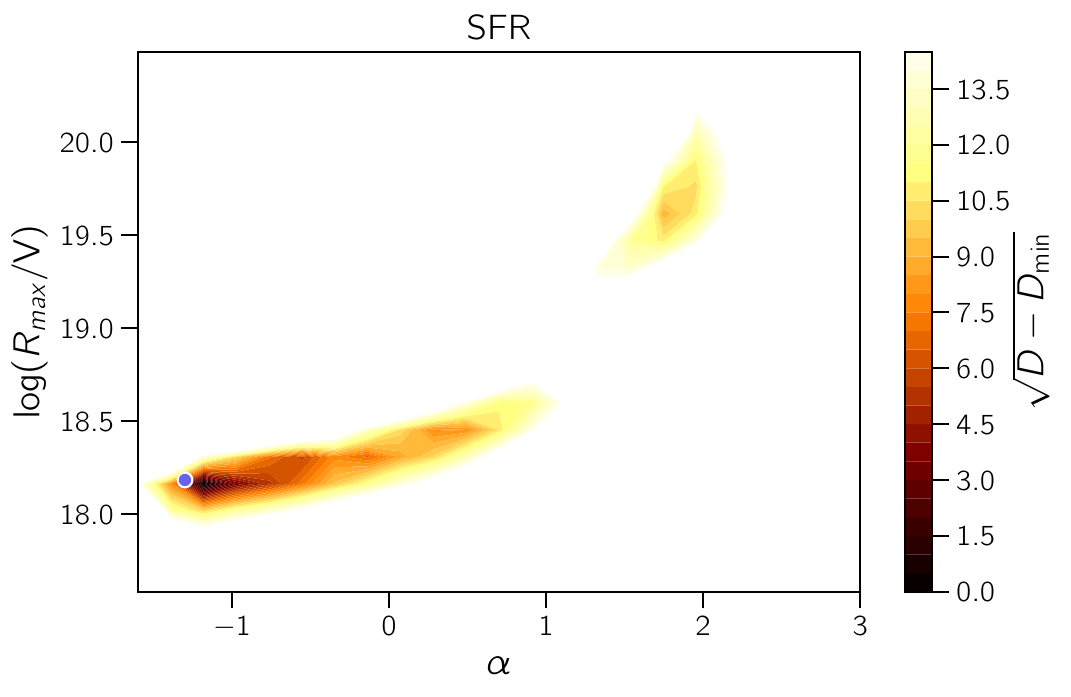}
  \includegraphics[width=.495\columnwidth]{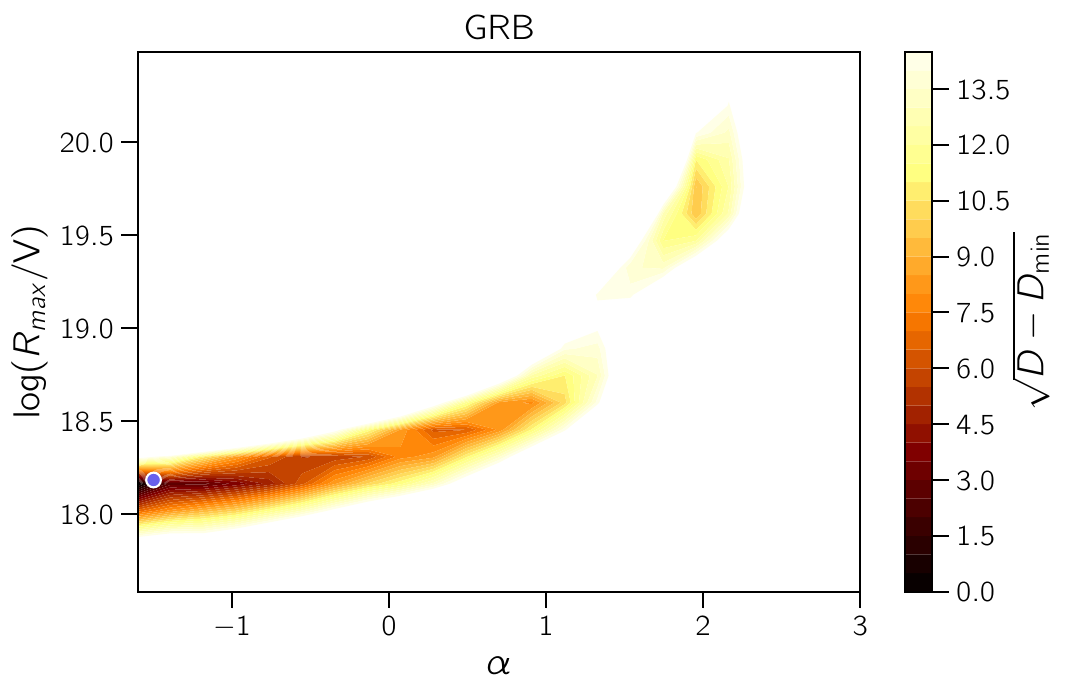}
  \includegraphics[width=.495\columnwidth]{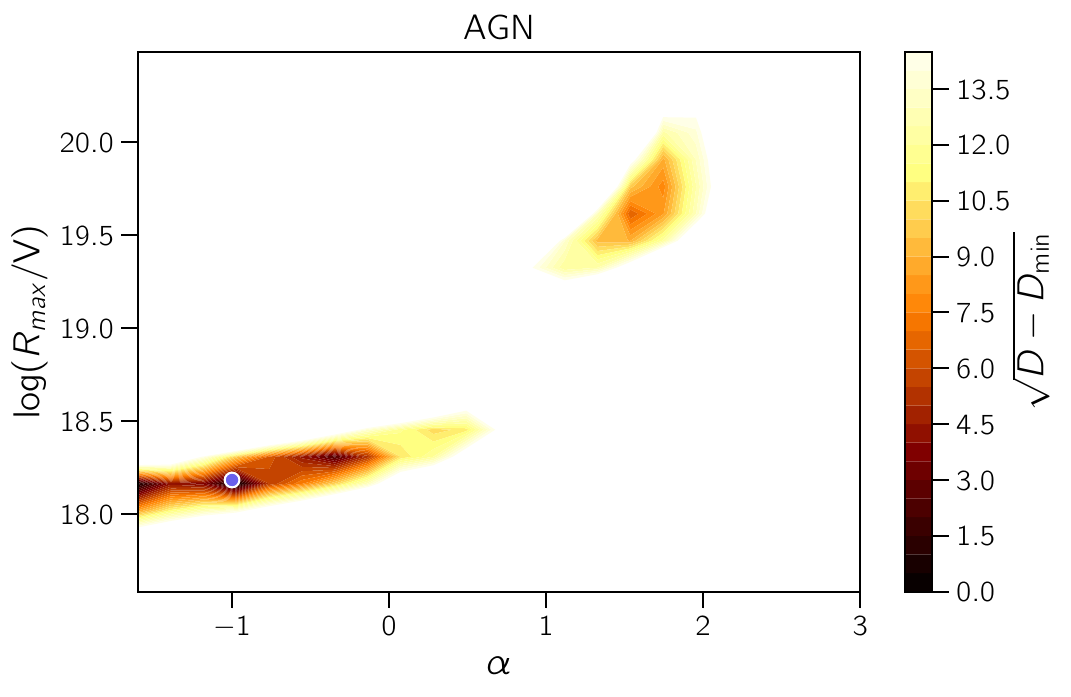}
  \caption{Parameter space of spectral index ($\alpha$) and maximal rigidity ($R_\text{max}$) for source evolution $m=-1.5$ (upper left), $m=+3.0$ (upper right), and $m=0$ (middle left panel); three cases of source evolutions are also shown, namely SFR (middle right), GRB (lower left), and AGN (lower right panel). The colour scale corresponds to the $\sqrt{D-D_\text{min}}$. The circle indicates the best-fit parameters.}
  \label{fig:bestFit_m}
\end{figure}

We now fix the source evolution in order to obtain the best-fit spectral index ($\alpha$) and maximal rigidity ($R_\text{max}$). This is illustrated in Fig.~\ref{fig:bestFit_m} for the cases of $m=-1.5$, $m=0$, $m=+3.0$, SFR, GRB, and AGN. 

\begin{figure}
  \centering
  \includegraphics[width=0.495\columnwidth]{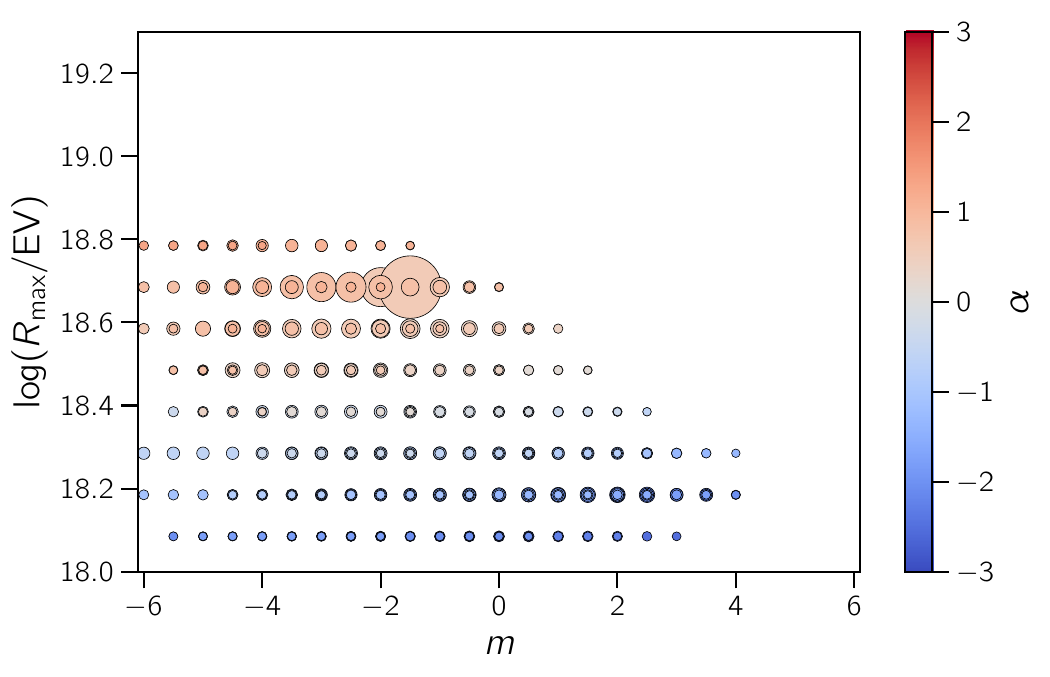}
  \includegraphics[width=0.495\columnwidth]{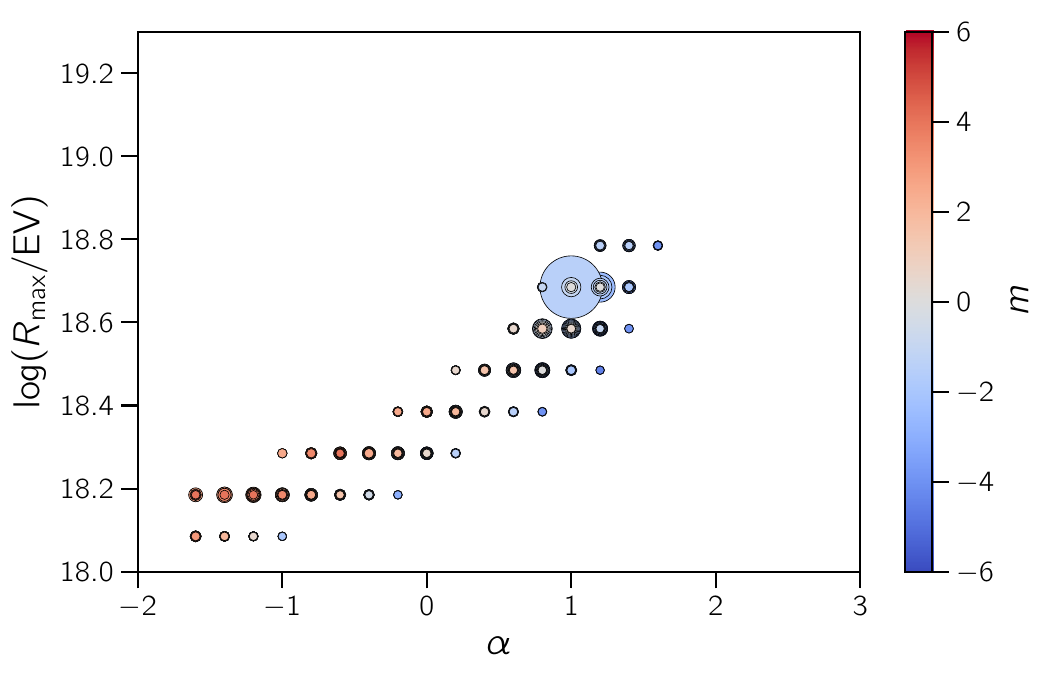}
  \caption{Best-fit values at 99\% confidence level for the maximal rigidity as a function of the source evolution for different spectral indices (left), and as function of the spectral index for different source evolutions (right panel). For reference, marker sizes are plotted with radii inversely proportional to the deviance of the corresponding scenarios. The colour scale corresponds to the spectral index (left panel) and to the source evolution parameter (right panel).}
  \label{fig:bestFitParameters}
\end{figure}

Figure~\ref{fig:bestFit_m} also suggests a trend that if source evolution is accounted for in the fit, then the spectral index tends to become increasingly larger for negative $m$. To study this dependence, we have compiled all pairs $(m, R_\text{max})$ that minimise $\sqrt{D - D_\text{min}}$  for a particular choice of $\alpha$; this is shown  in Fig.~\ref{fig:bestFitParameters}, left panel. Similarly, one can assume a fixed value for $m$ to obtain the values of $(\alpha, R_\text{max})$ that minimise the deviances, as shown in Fig.~\ref{fig:bestFitParameters}, right panel. In the left panel, there are few red markers in the region where $m > 0$, indicating that for positive source evolutions the best fit tends to favour negative spectral indices (and vice-versa). Similarly, in the right-hand-side panel one can see that for a fixed maximal rigidity, red circles (positive $m$) tend to be located in the region where $\alpha < 0$, whereas for $\alpha > 1$, negative source evolutions are preferred.

Our best-fit results are summarised in Table~\ref{tab:bestFit} for the complete analysis, as well as the specific cases of AGN, SFR, and GRB source evolutions. By computing the pseudo standard deviation, $\sqrt{D - D_\text{min}}$, we infer confidence intervals wherein the best-fit parameters $\alpha$, $R_\text{max}$, and $m$ would lie; this is shown in Table~\ref{tab:bestFitParameters}.

The choice of the pseudo standard deviation as an estimator is justfied within the approach we adopted. This follows Ref.~\cite{rolke2005a}. Our confidence intervals should be understood as the regions centred around the maximum likelihood estimator, limited by the curves corresponding to the desired percentile of a $\chi^2$-distribution with one degree of freedom. Because this is a multidimensional problem, the confidence regions need not be symmetric with respect to the corresponding best-fit parameters.

\begin{table}[h!]
  \centering
  \caption{Best-fit parameters for specific spectral indices.}
  \begin{tabular}{ccccccccccc}
  \hline
   $m$ & $\alpha$ & $\log(\frac{R_\text{max}}{\text{V}})$ & $f_\text{H}$ & $f_\text{He}$ & $f_\text{N}$ & $f_\text{Si}$ & $f_\text{Fe}$ & $\chi_\text{comp}^2$ & $\chi_\text{spec}^2$ & $\chi^2$ \\
  \hline
  -1.6 & +1.0 & 18.7 & 0.0003 & 0.0101 & 0.8906 & 0.0990 & 0.0 & 161.1 & 17.6 & 178.7 \\
   SFR & -1.3 & 18.2 & 0.1628 & 0.8046 & 0.0309 & 0.0018 & 0.0 & 184.4 & 19.9 & 204.3 \\
   AGN & -1.0 & 18.2 & 0.8716 & 0.0778 & 0.0469 & 0.0038 & 0.0 & 224.9 & 33.1 & 258.0 \\
   GRB & -1.5 & 18.2 & 0.5876 & 0.3973 & 0.0147 & 0.0004 & 0.0 & 170.8 & 20.1 & 190.9 \\
\end{tabular}
\label{tab:bestFit}
\end{table}

\begin{table}[h!]
  \centering
  \caption{Best-fit parameters for specific spectral indices.}
  \begin{tabular}{c|c|c|c|c}
  \hline
   C.L. & $D$ & \multicolumn{3}{c}{parameter range} \\
   \hline
    90\% & $< 2.71$ & $-4.3 \leq m \leq -0.7$ & $+0.8 \leq \alpha \leq +1.2$ & $18.6 \leq \log(R_\text{max} / \text{V}) \leq 18.7$ \\
    95\% & $< 3.84$ & $-5.5 \leq m \leq +3.5$ & $-1.6 \leq \alpha \leq +1.4$ & $18.2 \leq \log(R_\text{max} / \text{V}) \leq 18.7$ \\
    99\% & $< 6.63$ & $-6.0 \leq m \leq +4.2$ & $-1.6 \leq \alpha \leq +1.6$ & $18.1 \leq \log(R_\text{max} / \text{V}) \leq 18.8$ \\
    \hline
    90\% & $< 2.71$ & SFR & $-1.3 \leq \alpha \leq -1.2$ & $18.2 \leq \log(R_\text{max} / \text{V}) \leq 18.2$ \\
    95\% & $< 3.84$ & SFR & $-1.4 \leq \alpha \leq -0.6$ & $18.2 \leq \log(R_\text{max} / \text{V}) \leq 18.3$ \\
    99\% & $< 6.63$ & SFR & $-1.4 \leq \alpha \leq +0.1$ & $18.2 \leq \log(R_\text{max} / \text{V}) \leq 18.4$ \\
    \hline
    90\% & $< 2.71$ & AGN & $-1.0 \leq \alpha \leq -0.9$ & $18.2 \leq \log(R_\text{max} / \text{V}) \leq 18.2$ \\
    95\% & $< 3.84$ & AGN & $-1.0 \leq \alpha \leq -0.9$ & $18.2 \leq \log(R_\text{max} / \text{V}) \leq 18.2$ \\
    99\% & $< 6.63$ & AGN & $-1.0 \leq \alpha \leq -0.4$ & $18.2 \leq \log(R_\text{max} / \text{V}) \leq 18.3$ \\
    \hline
    90\% & $< 2.71$ & GRB & $-1.5 \leq \alpha \leq -1.4$ & $18.2 \leq \log(R_\text{max} / \text{V}) \leq 18.2$ \\
    95\% & $< 3.84$ & GRB & $-1.5 \leq \alpha \leq -0.7$ & $18.2 \leq \log(R_\text{max} / \text{V}) \leq 18.3$ \\
    99\% & $< 6.63$ & GRB & $-1.5 \leq \alpha \leq +0.4$ & $18.1 \leq \log(R_\text{max} / \text{V}) \leq 18.5$ \\
\end{tabular}
\label{tab:bestFitParameters}
\end{table}

Figure~\ref{fig:bestFitParameters} is instructive to constrain source models using the combined fit. In particular, for the most common spectral indices found in the literature ($1 \lesssim \alpha \lesssim 2.2$), scenarios with positive source evolution ($m > 0$) are  disfavoured. This confirms the results from Ref.~\cite{taylor2015a}. As a cautionary remark, one should bear in mind that a negative source evolution may be interpreted either as the actual evolution of sources (as in tidal disruption event populations~\cite{alvesbatista2017b,zhang2017a,biehl2017a,guepin2018a}) or as a mere representation of the dominance of nearby objects over distant ones, reflecting, to some extent, cosmic variance of the source distribution rather than a global behaviour. Evidence for a local overdensity of sources between $1 \times 10^{18}$ and $4 \times 10^{18} \; \text{eV}$ has been provided in Ref.~\cite{liu2016a}, and similar arguments may apply at higher energies.

From Table~\ref{tab:bestFitParameters}, it is hard to draw an conclusive results at confidence levels larger than 90\%. The spectral index, for instance, ranges from the lowest values considered, $\alpha = -1.6$, up to $\alpha=+1.4$ at 95\% C.L. Therefore, the constraining power of the fit is rather weak, and all our results should be interpreted with caution.

The relationship between source evolution and spectral index has also been discussed by the Auger Collaboration~\cite{auger2017a,wittkowski2017a}, confirming Ref.~\cite{taylor2015a}. In our approach we let the source evolution be a free parameter in the fit, using a fine spacing along this axis. This is important because cosmogenic fluxes strongly depend on the value of the source emissivity evolution, and a coarser spacing in $m$ could compromise the reliability of our predictions from Secs.~\ref{sec:cosmogenicResultsPh} and~\ref{sec:cosmogenicResultsNu}.


The scenarios SFR and AGN do not provide fits as good as the $(1+z)^m$ evolution, whereas the fit for the GRB is slightly better than for AGN and SFR, as indicated in Tab.~\ref{tab:bestFit}.

Interestingly, a local second minimum for $1.5 \lesssim \alpha \lesssim 2$ can also be seen in Fig.~\ref{fig:bestFit_m}. We have not investigated it separately. Nevertheless, it would allow for a much softer spectral index, compatible with common acceleration models, as well as trans-GZK protons. This minimum is more proncounced when the QGSJetII hadronic interaction model~\cite{ostapchenko2011a} is adopted, as noted in Ref.~\cite{auger2017a}.

The spectrum and the first two statistical moments of the $X_\text{max}$ distribution are shown in Fig.~\ref{fig:speccomp} for the $(1 + z)^m$ evolution. The best fits for the SFR, GRB, and AGN scenarios are shown in Appendix~\ref{app:fitRes}.
\begin{figure}
  \centering
  \includegraphics[width=0.50\columnwidth]{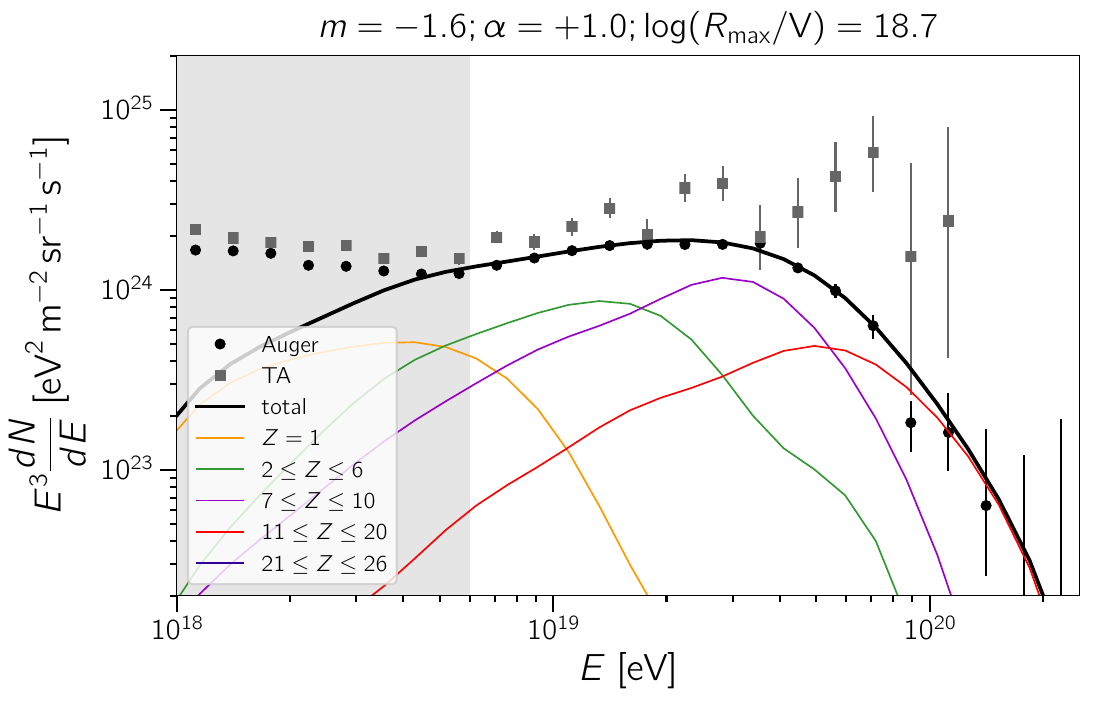} \\
  \includegraphics[width=0.495\columnwidth]{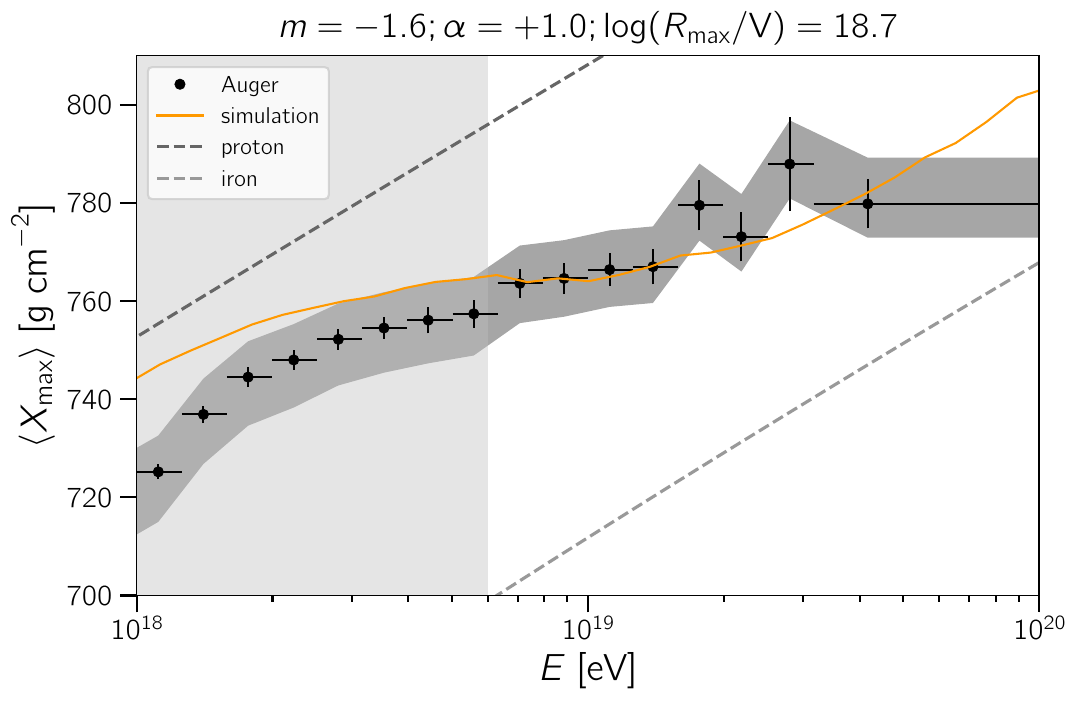}
  \includegraphics[width=0.495\columnwidth]{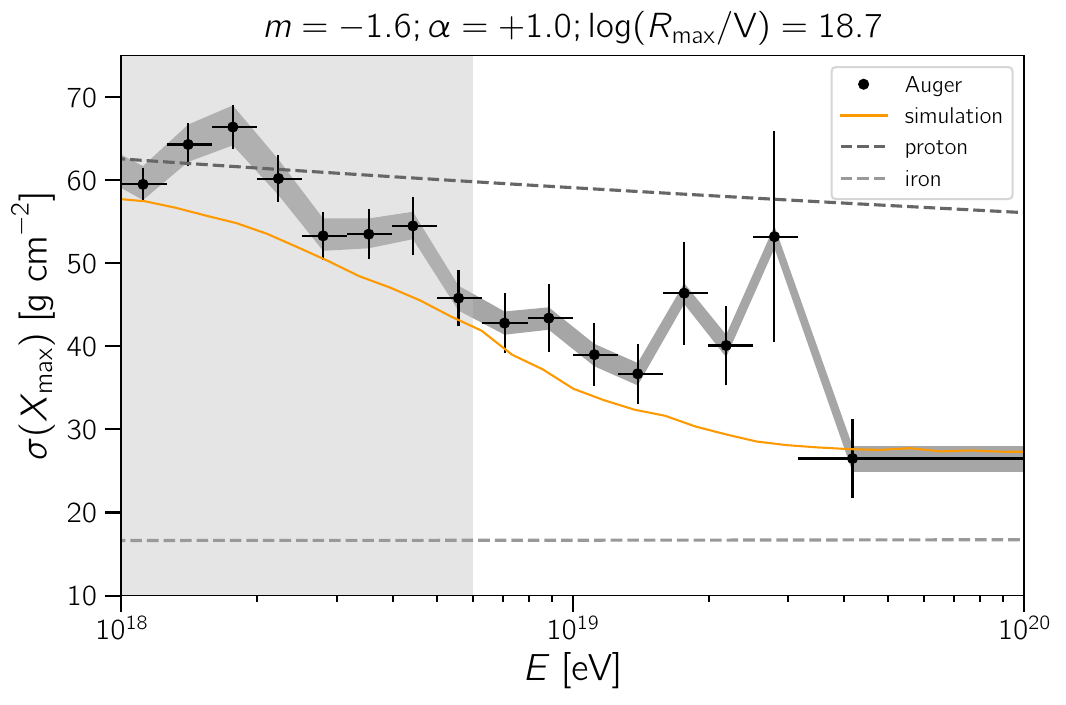}
  \caption{Spectrum (upper panel), $\langle X_\text{max} \rangle$ (lower left), and $\sigma(X_\text{max}$) (lower right panel) for the best fit to Auger data~\cite{auger2016b}. The spectrum measured by the Telescope Array~\cite{tinyakov2014a} is also shown for reference. The grey region at $E < 10^{18.7} \; \text{eV}$ refers to the energy range shown in the plots whose points were not used in fits. The dark grey regions around the composition-related observables correspond to the systematic uncertainties of Auger. The fractions of each injected element are: $f_\text{H}=0.0004$, $f_\text{He}=0.0002$, $f_\text{N}=0.8970$, $f_\text{Si}=0.1024$, and $f_\text{Fe}=0.0$.}
  \label{fig:speccomp}
\end{figure}

It is important to emphasise that we only consider $E > 10^{18.7} \; \text{eV}$. The best fit for the AGN case overshoots the measured UHECR spectrum for $E < 10^{18.7} \; \text{eV}$. Thus, this specific scenario can be ruled out. Nevertheless, because we have considered only medium-high-luminosity AGNs, the contribution of AGNs at other luminosity bands could change this picture, albeit high-luminosity AGNs evolve even more strongly ($m \simeq 7.1$) and the lower luminosity ones are likely not able to accelerate cosmic rays to ultra-high energies~\cite{lovelace1976a,waxman2004a}, as discussed in Sec.~\ref{sec:simulations}. One should also note that we have restricted ourselves to $\alpha \geq -1.6$, and the best fit could lie below this threshold. The best fits for the GRB and SFR scenarios also exceed the measured spectrum at some specific energies, but not as much as in the AGN case. In our phenomenological description we have used the same spectral index and maximal rigidity for all individual sources. In reality, these quantities are likely distributions, which may affect both the spectrum and composition.

We have assumed all sources to be equally luminous, which they are not. This may have a number of major consequences for our fit. An interesting approach was presented in Ref.~\cite{eichmann2017a}, who suggested that the maximal rigidity is related to the luminosity.

The fit is affected by the choice of hadronic interaction model. We have used only EPOS-LHC~\cite{pierog2013a}. As described in Ref.~\cite{auger2016a}, the deviances increase significantly when using the Sybill 2.1~\cite{ahn2009a} or QGSJetII-04~\cite{ostapchenko2011a} model. Interestingly, the latter model provides the best fit around $\alpha \simeq 2$, the second minimum seen in Fig.~\ref{fig:bestFit_m}.

Ref.~\cite{boncioli2016a} presents a comprehensive discussion on the effects of uncertainties in photonuclear cross sections for cosmic-ray propagation, providing a list of nuclides whose cross sections are poorly measured or not measured at all. Note that accurate photodisintegration cross sections are needed because they produce a cascade effect, such that uncertainties in the first interactions may incur large changes in the final results. One case in which the impact of this uncertainty is considerable relates to the ejection of $\alpha$-particles, which can result in a softer spectrum if intermediate-mass nuclei with hard spectra are injected~\cite{alvesbatista2016a}. This is particularly important in the context of our work as the best fit for the $(1 + z)^m$ evolution case is dominated by nitrogen ($f_\text{N} \approx 89\%$) with relatively hard spectrum ($\alpha = 1.0$).

A similar study has been recently performed by the authors of Ref.~\cite{romerowolf2018a}. They use bayesian methods to fit the UHECR spectrum considering a few specific source evolution models, obtaining a best fit for $\alpha=1.86$ and $\log(R_\text{max}/V)=18.3$, as well as a proton fraction of about 10\%. The main difference between our work and theirs is the prior distribution for $m$ and $\alpha$ used. In their case $-10 \leq \alpha \leq 10$. Moreover, their source evolution is different and does not encompass negative values of $m$. 

Little is known about extragalactic magnetic fields. For instance, intergalactic magnetic fields (IGMFs) occupy about $\sim 20 - 90\%$ of the total volume of the universe and their strength is estimated to lie in the range between $\sim 10^{-9} \; \text{G}$ and $\sim 10^{-17} \; \text{G}$ (for reviews see e.g. Refs.~\cite{kotera2011b,vallee2011a,durrer2013a}). 
The effects of magnetic fields on the spectrum and composition of UHECRs are not really well comprehended. The so-called propagation theorem states that, in the case of a uniform distribution of sources with separations much smaller than the typical propagation lengths, the UHECR spectrum has a universal shape independently of the modes of propagation~\cite{aloisio2004a}. Consequently, the assumption of a uniform source distribution is an adequate approximation in the limit of high source density, having magnetic fields little effect on the spectral shape. On the other hand, for relatively high magnetic fields the diffusion length may become comparable to the average source separation. In this limit, the propagation theorem no longer holds, thus affecting both the UHECR spectrum and composition~\cite{mollerach2013a}. Considering that voids dominate most of the volume of the known universe, more realistic magnetic field distribution could render unimportant such changes~\cite{alvesbatista2014a}.

The fit can be strongly affected by the distribution of sources and cosmic magnetic fields. In Ref.~\cite{wittkowski2017a} a similar study to this has been performed by the Pierre Auger Collaboration. More details about the simulations can be found in Ref.~\cite{wittkowski2018a}. A distribution of sources and magnetic fields obtained from a cosmological simulation of structure formation is used, yielding $\alpha = 1.61$ and $R_\text{max}=10^{18.88} \; \text{V}$ in the presence of intervening magnetic fields, as opposed to $\alpha=0.61$ and $R_\text{max}=10^{18.48} \; \text{V}$ in their absence. These results should be carefully interpreted nonetheless, because the fit has been proven to be sensitive to the distribution of matter and magnetic fields both of which are highly uncertain. Still, the conclusion that magnetic fields soften the best-fit spectrum in the fits holds true.

The lower limit on the density of UHECRs is $\sim 10^{-6} - 10^{-7} \; \text{Mpc}^{-3}$~\cite{auger2013b}. For equally luminous sources the UHECR flux scales roughly with the inverse of the distance squared, thus implying that nearby sources may dominate the flux. As a consequence, even if supposedly realistic matter distributions are used, unless it captures the distribution of UHECR sources in the local universe  accurately, major uncertainties may be introduced in the fit results. The assumption of a uniform source distribution, too, is likely not precise because it essentially neglects the spatial distribution of sources with respect to Earth as well as the granularity of such distribution, consequently affecting the distance to the closest (or most luminous) sources and hence the best-fit parameters.

\section{Cosmogenic photons}\label{sec:cosmogenicResultsPh}


The DGBR is the integrated contribution of unresolved sources and fluxes stemming from photon-producing processes such as interactions and decays. It has been measured with high precision by the Fermi Large Area Telescope (Fermi-LAT)~\cite{fermilat2015a}. Many components may contribute to the DGRB, including blazars~\cite{narumoto2006a,pavlidou2008a}, misaligned AGNs~\cite{dimauro2014a}, quasar outflows~\cite{wang2016a,wang2017a,liu2018a}, star-forming regions~\cite{fields2010a,makiya2011a,tamborra2014a}, the decay or annihilation  of very- and super-heavy dark matter~\cite{murase2012a,blanco2018a}, among others. For a detailed review the reader can refer to Ref.~\cite{fornasa2015a}. 

Cosmogenic photons are also a guaranteed contribution to the DGRB, although to which extent, it is uncertain. If UHECRs were purely protons, then a considerable flux would be expected, including UHE photons, depending on the distance to the nearest sources. If UHECRs were predominantly heavy nuclei, or if they had a mixed composition as the data seem to suggest, then a non-zero flux of cosmogenic photons would be expected due to inverse Compton scattering of nucleus-produced electrons (via Bethe-Heitler pair production), as well as nuclear decay and photopion production initiated by protons, provided that kinematic constraints were satisfied. Therefore, cosmogenic photons can be used to probe UHECR properties~\cite{hooper2011a,berezinsky2016a,supanitsky2016a,vanvliet2017a,vanvliet2017b,globus2017a}.

\begin{figure}
  \centering
  \includegraphics[width=0.5\columnwidth]{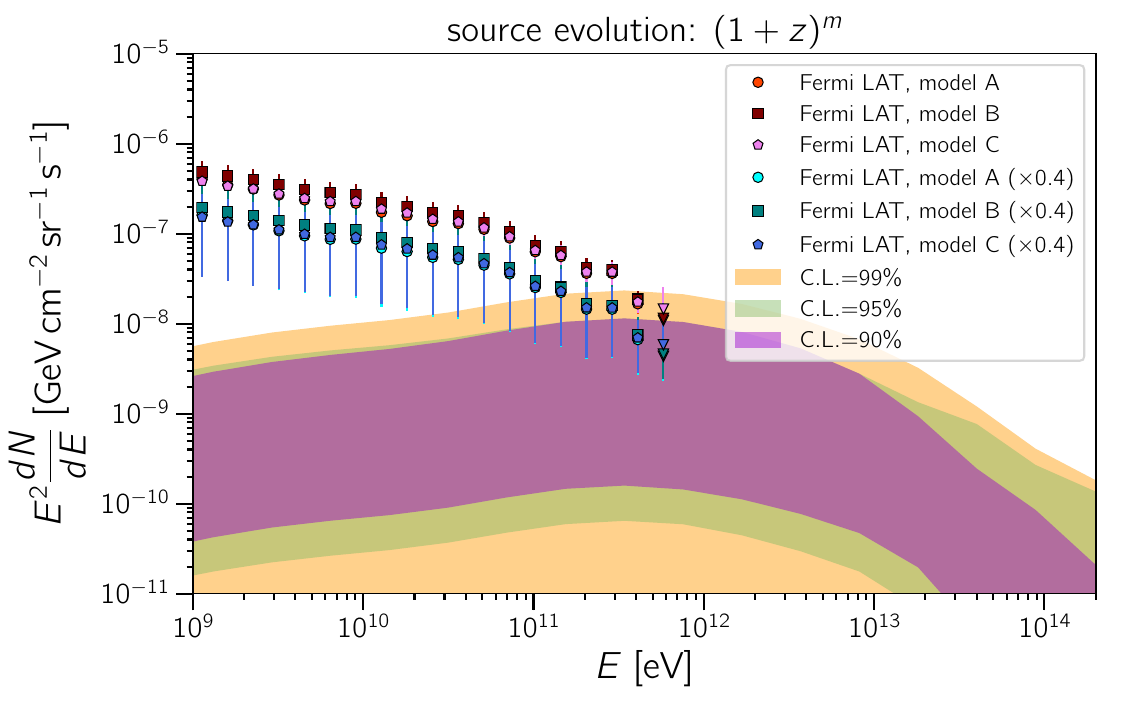}
  \caption{Cosmogenic photons for the best-fit scenarios with 90\%, 95\%, and 99\% confidence level, assuming a conservative cut on the flux beyond $z_\text{max}=1$. Fermi-LAT predictions for the DGRB~\cite{fermilat2015a} are also plotted for different galactic foreground models. The data is also shown scaled down by a factor 0.4 to account for possible unresolved point sources~\cite{fermilat2016a,lisanti2016a}. }
  \label{fig:cosmogenicPhotons}
\end{figure}

UHECR interactions and decays result in electrons and photons, which initiate electromagnetic cascades in the intergalactic medium. Charged particles produce synchrotron photons in the presence of magnetic fields. Intervening magnetic fields deplete part of the charged component of the cascade. For this reason, we have analysed two extreme cases for the IGMF, $B=0$ and $B=1 \; \text{nG}$, the latter being roughly the upper limit estimated by the Planck satellite~\cite{planck2016a}. Nevertheless, for $B=1 \; \text{nG}$ the spectrum is only affected by the magnetic field at energies $E \lesssim 10^{10} \; \text{eV}$. For this reason in Fig.~\ref{fig:cosmogenicPhotons} we show the gamma-ray spectrum for the case $B=0$ only, and compare it with Fermi measurements.

Fig.~\ref{fig:cosmogenicPhotons} shows the flux up to $E = 10^{15} \; \text{eV}$. Recent study by the Fermi Collaboration~\cite{fermilat2016a} suggests that point sources amount for $86^{+16}_{-14}\%$ of the DGRB. Ref.~\cite{lisanti2016a} reached similar conclusions. This implies that the DGRB can be conservatively scaled by a factor $\approx 0.4$, as shown in the figure. Consequently, a larger part of the parameter space shown in the figure can be constrained with Fermi-LAT data. 

The contribution of cascade photons to the DGRB is not well known, but it is possible to infer many of its properties. On a side note, AGNs can emit gamma rays with energies $E_\gamma \gtrsim 10 \; \text{TeV}$. Because in the standard AGN paradigm~\cite{urry1995a} blazar jets are approximately pointing towards Earth, this offers the possibility to constrain the strength of magnetic fields by studying the cascade component separately and observing changes in the low-energy region of the spectrum~\cite{alvesbatista2017c}, even in the more realistic case in which the jets are misaligned~\cite{gate2017a}. The upcoming Cherenkov Telescope Array (CTA)~\cite{cta2013a,cta2017a} will be able to thoroughly survey the sky and measure the DGRB with unprecedented precision. It is worth mentioning that UHECR-induced cascades are sometimes invoked to explain the hard spectra of some extreme TeV blazars with unusually hard spectra~\cite{essey2010a,essey2010b,essey2011a,aharonian2013a,essey2014a}. If this turns out to be the case, next-generation imaging air Cherenkov telescopes (IACTs) such as CTA might be able to resolve these objects, providing an adequate template to infer the actual contribution of this class of objects to the DGRB, thereby improving the UHECRs constrains we could derive. Although CTA may, in principle, be able to directly survey the full sky and measure the DGRB at $E \gtrsim 100 \; \text{GeV}$, this would be extremely difficult because a thorough understanding of the electron background would be necessary~\cite{sol2013a}.

Our results from Fig.~\ref{fig:cosmogenicPhotons} are compatible with Ref.~\cite{liu2016a}, which suggests that the highest-energy bin of Fermi-LAT at 820 GeV favours the existence of a local overdensity of UHECRs, or a higher local energy injection rate compared to that of distant sources, dominating the cosmic-ray spectrum at sub-ankle energies of $E \lesssim 10^{18.7} \; \text{eV}$. The fact that our fit favours negative source evolution is in consonance with this explanation. 


None of our best-fit bands exceed the measurements of the DGRB measured by Fermi if error bars and model uncertainties are included. If these error bars are ignored, for some DGRB models, especially the ones scaled down by a factor 0.4, some of the UHECR parameter space could already be (weakly) constrained, in particular combinations of large $m$, large proton fraction at the source, and low spectral indices. This confirms the results of Ref.~\cite{vanvliet2017a}.

We have set $z_\text{max} = 1$. For larger $z_\text{max}$, the photon fluxes could change,  as discussed in Appendix~\ref{app:zmax}. As a consequence, if UHECR sources are distributed up to $z_\text{max} \approx 5$, for instance, then they could represent a sizeable fraction of the total DGRB, severely constraining the combined contribution of other DGRB components. Stronger source evolutions ($m > 0$) would lead to even higher fluxes. 

At the highest energies, the flux of photons is very small. In the energy range  $10^{18} - 10^{20} \; \text{eV}$, limits by Auger are $E_\gamma ^2 \Phi_\gamma \sim 3-12 \times10^{-7} \; \text{GeV} \, \text{cm}^{-2} \, \text{s}^{-1}$, at 95\%~C.L.~\cite{bleve2015a,auger2017d}. At $E \sim 10^{20} \; \text{eV}$, limits by TA are $E_\gamma ^2 \Phi_\gamma \sim 2 \times10^{-6} \; \text{GeV} \, \text{cm}^{-2} \, \text{s}^{-1}$~\cite{rubtsov2015a}. Above $10^{18} \; \text{eV}$, our photon fluxes are $E_\gamma ^2 \Phi_\gamma \lesssim 10^{-14} \; \text{GeV} \, \text{cm}^{-2} \, \text{s}^{-1}$. As a consequence, for low-rigidity scenarios with low proton content, the prospects for detecting UHE photons seem bleak, given that the expected fluxes are several orders of magnitude below the sensitivity of current- and next-generation experiments.

\section{Cosmogenic neutrinos}\label{sec:cosmogenicResultsNu}

\begin{figure}
  \centering
  \includegraphics[width=0.495\columnwidth]{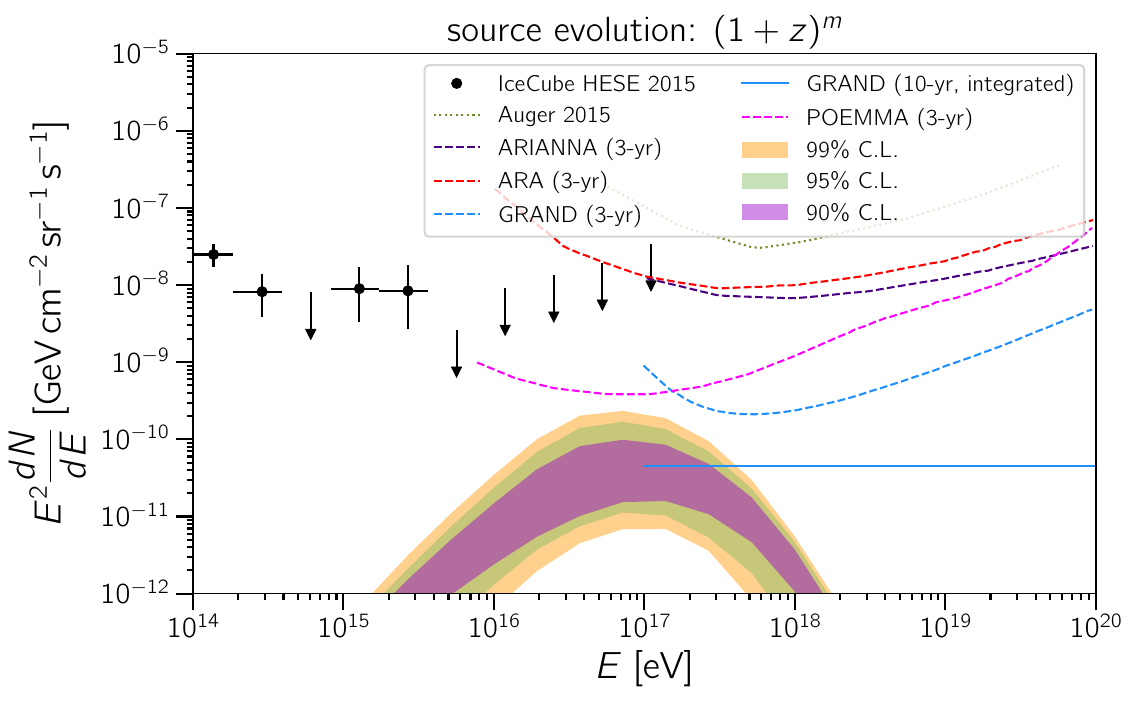}
  \includegraphics[width=0.495\columnwidth]{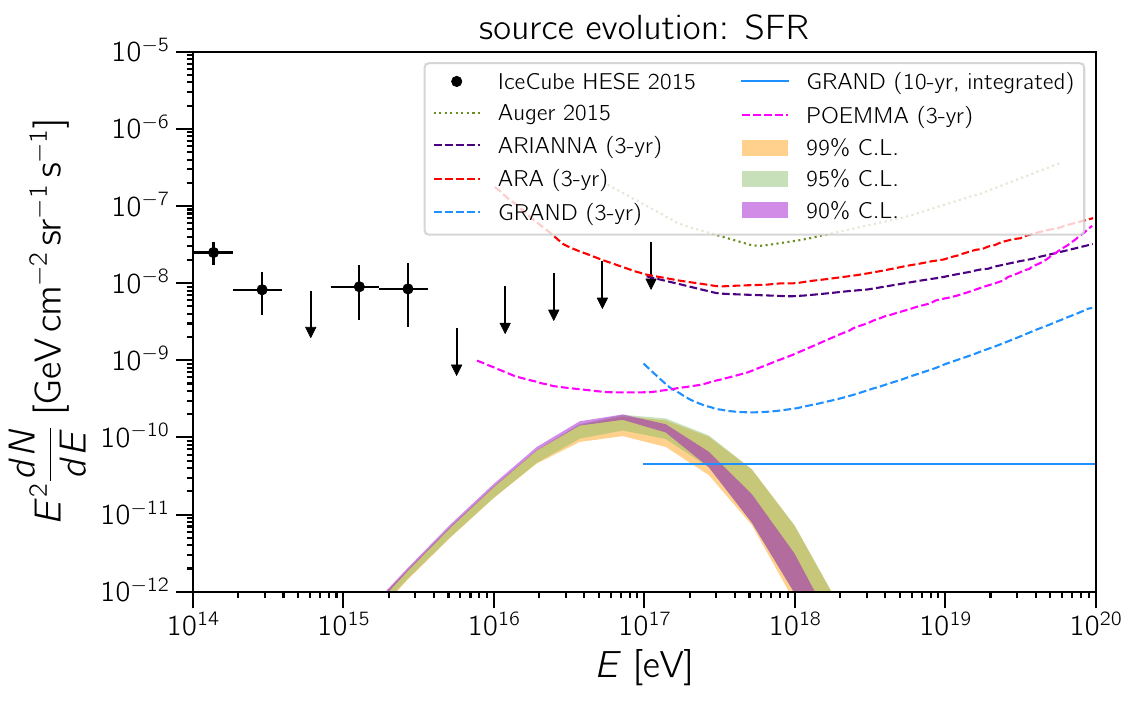}
  \includegraphics[width=0.495\columnwidth]{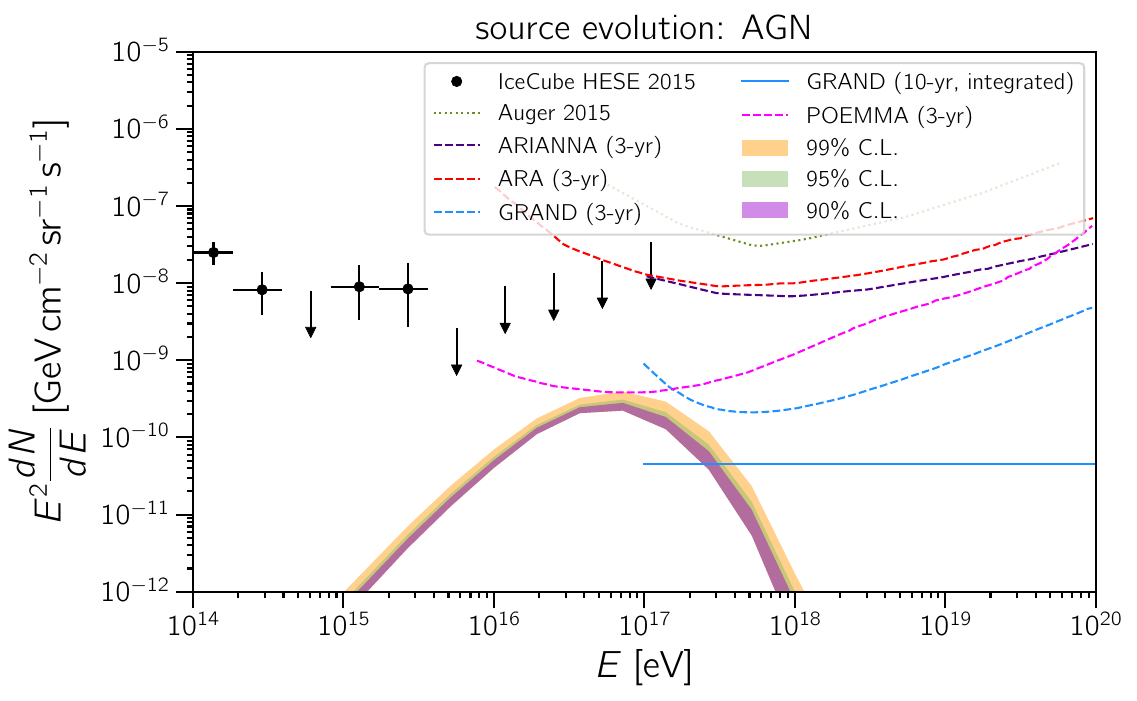}
  \includegraphics[width=0.495\columnwidth]{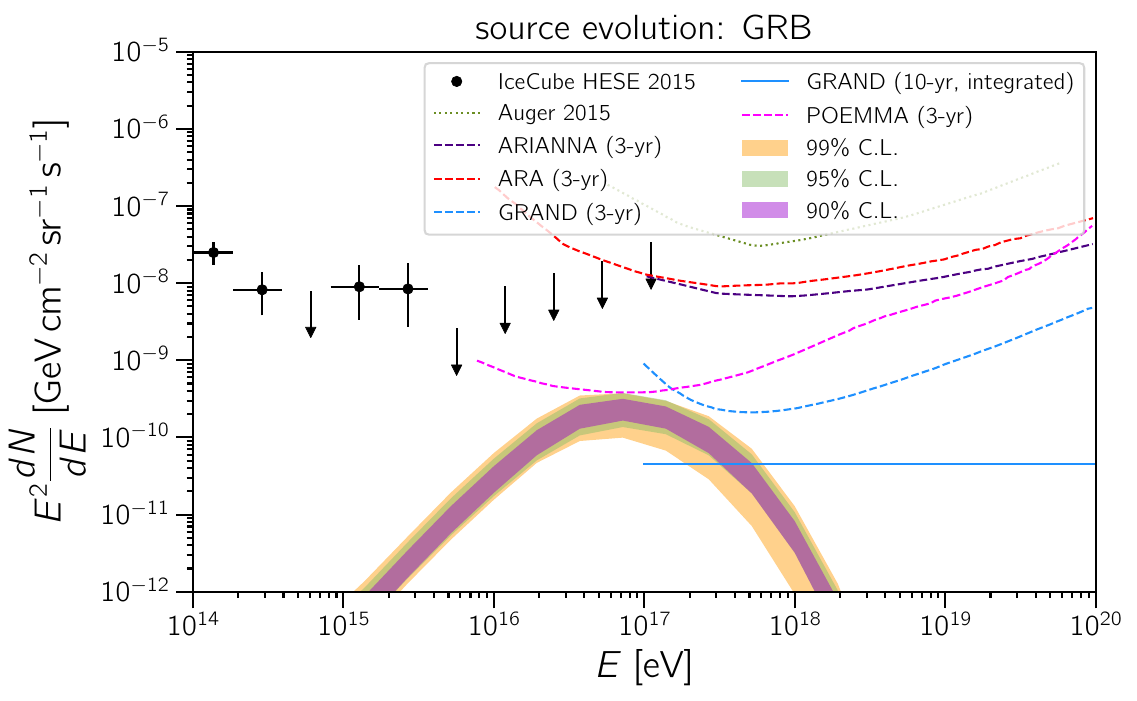}
  \caption{All-flavour ($\nu_e : \nu_\mu : \nu_\tau = 1:1:1$) cosmogenic neutrino fluxes for the best-fit scenarios with 90\%, 95\%, and 99\% confidence level. The sensitivity curves for Auger~\cite{auger2015a} (dotted lines) is shown, together with IceCube HESE events~\cite{icecube2015a} (black circles). The projected 3-year sensitivities for ARIANNA~\cite{arianna2015a}, ARA~\cite{ara2012a}, POEMMA~\cite{poemma2017a}, and GRAND~\cite{grand2017a,grand2017b,grand2018a} are also displayed as dashed lines. The upper left panel corresponds to a source evolution $(1 + z)^m$, the upper right to SFR, and the left and right lower rows are for AGN and GRB evolutions, respectively, assuming a conservative cut on the flux beyond $z_\text{max}=1$. The predictions for more realistic cuts ($z_\text{max} = 5$) are presented in Appendix~\ref{app:zmax}.}
  \label{fig:cosmogenicNeutrinos}
\end{figure}

Similarly to Sec.~\ref{sec:cosmogenicResultsPh}, we have also calculated the neutrino spectrum for the best-fit scenarios, as shown in Fig.~\ref{fig:cosmogenicNeutrinos}. We study the following evolutions: SFR, AGN, GRB, and $(1+z)^m$. Unlike the photon flux which is essentially just scaled up or down depending on the composition, injection spectrum, and source evolution model, in the case of neutrinos the spectrum changes significantly due to the absence of a horizon, as shown in Ref.~\cite{vanvliet2017b}.

The neutrino spectrum for the best-fit scenarios are shown in Fig.~\ref{fig:cosmogenicNeutrinos}. We represent all the scenarios through bands encompassing the limiting cases at 90\%, 95\%, and 99\% confidence levels. The bands for the $(1+z)^m$ scenario include the low-rigidity iron-rich estimate from Ref.~\cite{kotera2010a}, for which $E_\nu ^2 \Phi_\nu \sim  10^{-10} \; \text{GeV} \, \text{cm}^{-2} \, \text{s}^{-1}$ at $E \approx 3 \times 10^{17} \; \text{eV}$. The factor of a few difference in level stems mainly from the harder injection spectrum and the negative source evolution derived from our fits.

We overlay in Figure~\ref{fig:cosmogenicNeutrinos} the IceCube HESE events~\cite{icecube2015a}, together with the sensitivity curves for Auger~\cite{auger2015a} and the projected 3-year sensitivities for ARIANNA~\cite{arianna2015a}, ARA~\cite{ara2012a}, POEMMA~\cite{poemma2017a}, and GRAND~\cite{grand2017a,grand2017b,grand2018a}. For reference, we also show the 10-year integrated sensitivity of GRAND, corresponding to a null signal detection, assuming a background-free scenario. This upper limit is obtained by requiring an maximum number of 2.44 events, integrating over a neutrino flux following a power-law in $E_\nu^{-2}$  \cite{feldman1997a}. The level of this upper limit is low enough to ensure that the most conservative scenarios will be probed with future instruments, if they reach their projected sensitivities. 

Note that the specific evolution scenarios SFR, GRB, and AGN provide relatively high cosmogenic fluxes, compared to the $(1 + z)^m$ case, despite their slightly higher deviances, as shown in Tab.~\ref{tab:bestFit}. Therefore, the prospects for detecting cosmogenic neutrinos are good if the sources evolve as SFR, GRB, or AGN, which comprise many UHECR source candidates found in the literature.

All our estimates are conservative as we do not account for neutrinos coming from $z > 1$. For $m < 0$, this assumption has little impact on the results; however, for $m > 0$, which encompasses the AGN, GRB, and SFR cases, the fluxes of cosmogenic neutrinos could increase significantly. Therefore, ours fluxes are deliberately underestimated. The impact of the choice of $z_\text{max}$ on the cosmogenic spectra is generally discussed in Appendix~\ref{app:zmax}. We also present predictions of the fluxes for the SFR, AGN, and GRB scenarios for $z_\text{max} = 5$. 

Our results are compatible with the upper limit estimated by the Pierre Auger Observatory~\cite{auger2015a} for a single-flavour $E^{-2}_\nu$ flux, which is $E_\nu^2 dN_\nu / dE \lesssim 6 \times 10^{-9} \; \text{GeV} \, \text{cm}^{-2} \, \text{s}^{-1}$ in the energy range between $10^{17} \; \text{eV}$ and $2.5 \times 10^{19} \; \text{eV}$. If UHE neutrinos are detected by Auger in this energy range, with fluxes slightly below the quoted limits, then they would likely have been produced {\it in situ}, via UHECR interactions with the matter and radiation fields surrounding the sources.

We have used the EBL model by Gilmore et al.~\cite{gilmore2012a}. The impact of different EBL models on the neutrino spectrum is less than a few percent at $E \gtrsim 10^{18} \; \text{eV}$, but it can be high ($\gtrsim 10\%$) at $E \lesssim 10^{17} \; \text{eV}$. In this same energy range, uncertainties in photonuclear cross sections can also affect the spectrum, introducing discrepancies of up to $\sim 30\%$ for hard nitrogen injection~\cite{alvesbatista2018a}. 

A recent study by Romero-Wolf \& Ave~\cite{romerowolf2018a} estimate the cosmogenic neutrino flux using bayesian methods to infer the best-fit parameters, obtaining $E_\nu ^2 \Phi_\nu \sim 10^{-11} \; \text{GeV} \, \text{cm}^{-2} \, \text{s}^{-1} \, \text{sr}^{-1}$, which is roughly consistent with our results. The actual best-fit parameters are, however, significantly different from ours. One reason for that is the fact that they consider a wider range of parameters. Also, they have used a more detailed source evolution model composed of three parts: the first grows with redshift, reaching a plateau at intermediate redshifts, and then exponentially decaying until it is completely suppressed, vaguely resembling our scenarios for AGN, SFR, and GRB. Neverthless, by enforcing that the density of sources increase with redshift between redshifts $z=0$ and $z_1 > 0$, the overall behaviour of a negative source evolution is not properly captured. As a consequence, the mean source evolution is necessarily positive and the best-fit spectral index is shifted towards lower values, as shown in Fig.~\ref{fig:bestFitParameters}. 
Although a negative injection spectral index would significantly decrease the flux of cosmogenic neutrinos, this is compensated by the source evolution assumptions made by the authors of Ref.~\cite{romerowolf2018a}. Therefore, it is consistent that their best-fit result lies within the uncertainty band shown in Fig.~\ref{fig:cosmogenicNeutrinos}.

ANITA-I~\cite{anita2016a} has reported a possible upcoming event with $E = (6 \pm 4) \times 10^{17} \; \text{eV}$. This could be a genuine ultra-high energy event, but neither Auger nor IceCube have detected similar events, as it would be expected from a diffuse flux. Another possibility is a transient event. Because the $\nu_\tau$ would be absorbed by the ice, this event defies conventional explanations, leaving room for a beyond Standard Model (BSM) explanation to justify its detection~\cite{romerowolf2017a}.

Results by ANITA-III~\cite{anita2018a} provide evidence for a possible event with energy $E > 10^{19} \; \text{eV}$. Although compatible with the background, the hypothesis that the observed event is a genuine UHE neutrino withstood further scrutiny. In light of the relatively low cosmogenic neutrino fluxes we have computed, if the ANITA-III event is genuinely of astrophysical origin, then it should stem from a luminous neutrino source greatly exceeding the flux of cosmogenic neutrinos. Another possibility is to invoke a BSM explanation to justify its detection, similarly to the possible event detected by ANITA-I.

\section{Conclusion and Outlook}\label{sec:conclusions}

The latest results from the Pierre Auger Observatory constrain some of the key parameters of ultra-high-energy cosmic-ray source models. By scanning these parameters, namely, the source emissivity evolution history, spectral index at injection, maximal rigidity and chemical composition, we obtain best-fit regions for which we compute the associated cosmogenic neutrino and photon fluxes produced by particles during their propagation in the intergalactic medium. 

In this analysis we have considered the following emissivity evolutions: SFR, AGN, GRB, and $(1 + z)^m$. The two latter scenarios lead to similar levels of deviances, although the $(1 + z)^m$ model results in slightly better fits. Negative spectral indices are obtained for specific source evolutions: $\alpha = -1.3$ for SFR, $\alpha = -1.0$ for AGN, and $\alpha = -1.5$ for GRB, all of them with very low maximal rigidities ($\log(R_\text{max} / \text{V}) = 18.2$).

If we assume the emissivity evolution to be a free parameter, our best fit is obtained for $\alpha \simeq 1$, for compositions at injection dominated by intermediate-mass nuclei (nitrogen and silicon groups). The value of $m$ that minimises the deviance is $m=-1.6$. Evolutions with $m < -0.7$ are favoured at a 90\% C.L.. A negative source evolution may occur in certain populations of sources such as tidal disruption events. Another possible explanation is cosmic variance, in which case the local distribution of sources can mimic a negative evolution, while in reality there may be one or more dominant sources with an overall positive (e.g. star formation type) or null source evolution. Low maximal rigidities are natural, and the injection of intermediate-mass nuclei can be justified within the context of models involving acceleration in regions with significant metallicities, such as stars. Furthermore, there is a degeneracy between source number density and luminosity evolutions, such that the real redshift dependence should take into account both of these quantities. 

As discussed in the introduction, the hard spectral indices are difficult to be reconciled with most particle acceleration models. Although there are some models in the literature that predict the production of diracs at some energies and others that involve enhanced escape of higher energy particles, a further softening will happen due to the distribution of parameters across the source population. Note that radiation and magnetic fields in the source environment may affect the escape of UHECRs from sources, effectively hardening or softening the spectrum.

Our best fits for $(1 + z)^m$ with the hardest spectral indices provide the lowest possible cosmogenic fluxes compatible with the observed data, given our assumption, and can thus be viewed as \emph{pessimistic} scenarios. Note that these predictions depend on the maximal redshift, which we have set to 1; in reality, $z_\text{max}$ can be much higher, consequently increasing the cosmogenic fluxes. Therefore, our claim that ours is the most conservative cosmogenic fluxes compatible with Auger data is fully justified and overly pessimistic. 

For photons, the flux levels are in agreement with previously calculated fluxes~\cite{decerprit2011a,vanvliet2017a,vanvliet2017b}. Some of the scenarios nearly exceed the Fermi measurements of the DGRB. This sets strong bounds on the DGRB at energies $\sim 580-820 \; \text{GeV}$ because some UHECR models would leave little to no room for the contribution of specific source populations. This also constrains models that predict a significant contribution of dark matter decay and annihillation in this energy range.

For neutrinos, we find fluxes as low as $E_\nu ^2 \Phi_\nu \sim 4 \times 10^{-12} \; \text{GeV} \, \text{cm}^{-2} \, \text{s}^{-1} \, \text{sr}^{-1}$ at $E \approx 10^{17} \; \text{eV}$. This is  below previously predicted lower limits~\cite{kotera2010a}, due to the relaxed assumption on the spectral index and choice of $z_\text{max}$. At $E \sim 10^{17} \; \text{eV}$, the most pessimistic neutrino fluxes for the SFR, AGN, and GRB scenarios are, respectively, $1 \times 10^{-10} \; \text{GeV} \, \text{cm}^{-2} \, \text{s}^{-1} \, \text{sr}^{-1}$, $ 2 \times 10^{-9} \; \text{GeV} \, \text{cm}^{-2} \, \text{s}^{-1} \, \text{sr}^{-1}$, and $3 \times 10^{-11} \; \text{GeV} \, \text{cm}^{-2} \, \text{s}^{-1} \, \text{sr}^{-1}$. 
If we consider the more realistic scenarios with the contribution of sources at high redshifts, the cosmogenic fluxes for the SFR, AGN, and GRB scenario would range between $ 8 \times 10^{-10}$ and $ 2 \times 10^{-9} \; \text{GeV} \, \text{cm}^{-2} \, \text{s}^{-1} \, \text{sr}^{-1}$, for $z_\text{max} = 5$. If we treat the source evolution as a free parameter, the obtained fluxes are low even if we relax the constraint on $z_\text{max}$, due to the degeneracy between $\alpha$ and $m$ shown in Fig.~\ref{fig:bestFitParameters}.

The constraining power of the fit is rather weak, and most of the parameter space cannot be constrained at confidence levels larger than 90\% C.L.. A way to improve this picture would be to add other ingredients to the model. For instance, spectral indices and maximal rigidities could be treated as distributions rather than single values. Other source distributions with non-uniform luminosities could be used as well. A more detailed treatment would include other hadronic interaction models, uncertainties on the EBL~\cite{alvesbatista2015a,alvesbatista2018a}, as well as uncertainties stemming from photodisintegration~\cite{alvesbatista2015a,boncioli2016a,soriano2018a}.

As discussed in section~\ref{sec:fitResults}, taking into account extragalactic magnetic fields could soften the propagated spectrum and lead to increased production of secondary particles, thereby enhancing the fluxes of cosmogenic photons and neutrinos. Yet, this magnetic horizon effect is highly dependent on the density and distribution of sources, as well as on the distribution and properties of magnetic fields~\cite{alvesbatista2017a}.

Figure~\ref{fig:cosmogenicNeutrinos} shows that part of the conservative fluxes derived in this study are within reach of two projected experiments: GRAND and POEMMA. GRAND could detect some events with $3-10$ years of operation. Combined with POEMMA, whose sensitivity dips at around $0.1 \; \text{EeV}$ (whereas GRAND is at $\sim 0.5 \; \text{EeV}$) it would nicely constrain most models due to the position of the cosmogenic neutrino bumps. 

Our cosmogenic neutrino fluxes are rather low, due to the choice of $z_\text{max}=1$. Nevertheless, even in this pessimistic case, GRAND would be able to detect cosmogenic neutrinos for sources evolving as SFR, GRB, and AGN, at a 99\% C.L., within the first 5 years of operation. The sensitivities of ARA and ARIANNA are a few orders of magnitude below those of GRAND and POEMMA. Nevertheless, specific source populations with SFR, GRB, and AGN evolutions could be probed by relaxing the excessively strict constrain of $z_\text{max}=1$, as shown in Appendix~\ref{app:zmax}. If sources evolve as $(1 + z)^m$, the detection of cosmogenic neutrinos would be possible but not certain.

Our study demonstrates that the detection of cosmogenic neutrinos in the worst-case scenarios is favoured but not guaranteed, provided that the projected instruments reach their expected sensitivities and that they operate for over a decade.
From a different perspective, low cosmogenic fluxes could be profitable for EeV neutrino astronomy. Such a scenario would imply that the neutrinos first detected by future experiments would likely be those produced directly at the sources, via interactions of UHECRs with photon and baryon fields in the source environment. Abundant interactions should happen at the acceleration site of UHECRs, and theoretical models predict fluxes that are much higher than the level of cosmogenic neutrinos estimated there. In that case, it is advantageous that the cosmogenic neutrinos would constitute a low-level background, easing the identification of the first UHE neutrino point sources.

\acknowledgments

We thank Mauricio Bustamante, Kohta Murase, and Michael Unger for useful discussions. We also thank the Armando di Matteo and the Pierre Auger Collaboration for providing the systematic uncertainties used in the fit. We greatly appreciate the detailed feedback and cross checks performed by Jonas Heinze. We would also like to thank the anonymous referee for carefully checking our results and for the comments that helped us improve this manuscript.

RAB is supported by grant \#2017/12828-4, São Paulo Research Foundation (FAPESP). 
The work of RMdA and BL were partially supported by the Conselho Nacional de Desenvolvimento Cientifico e Tecnológico (CNPq) and by the Fundação Carlos Chagas Filho de Amparo à Pesquisa do Estado do Rio de Janeiro (FAPERJ).
KK is supported by the APACHE grant (ANR-16-CE31-0001) of the French Agence Nationale de la Recherche.

\bibliographystyle{JHEP}
\bibliography{references}

\appendix
\section{Fitting procedure}\label{app:fitting}

We performed a combined fit of the energy spectrum and mass composition measured by the Pierre Auger Observatory~\cite{auger2017a}. We have extended the space of parameters of this fit by adding the evolution of the source emissivity as a free parameter. The dataset fitted in this work is the same used in Ref.~\cite{auger2017a}. We have developed our analysis tools following the procedure described in this reference. We use the deviance ($D$) as proxy for the goodness-of-fit, which is defined as
\begin{equation}
    D = D(J) + D(X_\text{max})= -2 \ln \frac{L_J}{L_J^\text{sat}} - 2 \ln \frac{L_{X_\mathrm{max}}}{L_{X_\mathrm{max}}^\text{sat}},
\end{equation}
wherein $L_J$ and $L_{X_\mathrm{max}}$ are the likelihood functions of a given model, and $L_J^\text{sat}$ and $L_{X_\mathrm{max}}^\text{sat}$ are the corresponding ones of a saturated model that matches the data. A uniform scan over the spectral index $\alpha$, maximal rigidity $\log(R_\text{max} / \text{V})$, and source evolution index $m$, is performed. The deviance is minimised with respect to the relative abundances of the injected nuclei ($f_A$) using the Minuit package~\cite{james1975a}. 

The scan is performed in the intervals $\alpha=[-1.6,  3.0]$, $\log(R_\text{max}/\text{V})=[17.5,  20.5]$, and $m=[-6, 6]$ for $0 \leq z \leq 1$, on a grid with spacings of $0.1$ in $\alpha$,  $0.1$ in $\log_{10}(R_\text{cut}/\text{V})$ and $0.5$ in $m$. The results are then resampled onto a grid with spacing 0.1 in $m$. These intervals were chosen to encompass typical source models. Our dataset contains $10^{6}$ events for each nuclear species (H, He, N, Si, and Fe), which are reweighted. 

All the simulated events were binned in intervals of $\log(E / \text{eV}) = 0.1$ for direct comparison with the Auger data. The composition injected at the source is assumed to be uniform in the interval $[0, 1]$, and the condition $f_\text{H} + f_\text{He} + f_\text{N} + f_\text{Si} + f_\text{Fe} = 1$ is enforced.

The energy spectrum used in the fit~\cite{auger2016b} is composed by the sum of vertical and inclined events, comprising a total of 47767 surface detector events distributed in 15 bins of 0.1 in $\log(E / \text{eV})$ ($18.7 \leq \log(E / \text{eV}) \leq 20.2$). Since the Auger energy spectrum is an unbiased measurement of the flux corrected for the detector response, a forward folding procedure~\cite{auger2010b,auger2015c} was applied to each simulated spectrum before the comparison with the measurements. The deviance ($D$) is defined as the sum in each bin $i$ of $\log (E / \text{eV})$, and is given by
\begin{equation}	
   \label{eq:devsum}
    D = -2 \sum_i \mu_i - n_i + n_i \ln \left( \frac{n_i}{\mu_i} \right),
\end{equation}
where $n_i$ denotes the observed counts in the $i-$th (logarithmic) energy bin, and $\mu_i$ is the corresponding expected number of events obtained from the simulations.

The depth of maximum $(X_\text{max})$ distributions used in the fit~\cite{auger2014a} are composed of 1446 fluorescence detector events separated in bins of $20 \; \text{g}\,\text{cm}^{-2}$ in the same energy bins of the spectrum up to $\log(E / \text{eV}) = 19.5$, followed by a final from $\log(E / \text{eV}) = 19.5$ to $\log(E / \text{eV}) = 20.0$. The mass composition of the arriving simulated particles are inferred by using the parametrisation presented in Ref.~\cite{auger2017a}, which makes use of the Gumbel function $g(X_\text{max} | E, A)$. The Gumbel distributions are then corrected for detectors effects, as done in Ref.~\cite{auger2017a}, in order to obtain the expected model probability $(G_{i}^\text{model})$, evaluated at the logarithmic average of the energies of the observed events in the $i$-th bin. Therefore, the probability of observing a $X_\text{max}$ distribution $\vec{k}_i = (k_{i1}, k_{i2}, ...)$ is given by a multinomial distribution, which reads:
\begin{equation}
    L_{X_\mathrm{max}} = \prod\limits_i n_i! \prod \limits_x \frac{1}{k_{i x}!} (G_{i x}^\text{model})^{k_{i x}},
    \label{eq:likelihood_xmax}
\end{equation}
where $G_{i x}^\text{model}$ is the probability to observe an event in the  $X_\text{max}$ bin $x$.

In the fit performed by the Auger Collaboration~\cite{auger2017a}, systematic uncertainties are taking into account. An uncertainty of about 14\% in the energy scale was considered. The systematics for the composition depend on the energy, but are less than $9 \; \text{g}\,\text{cm}^{-2}$ over the whole energy range. In our study we have neglected systematics because the uncertainties due to hadronic interactions, EBL, and photodisintegration cross sections have a larger effect on the neutrino and photon flux predictions. Detailed discussions on the impact of these uncertainties on the spectrum and composition can be found in Ref.~\cite{auger2017a,alvesbatista2015a,boncioli2016a,soriano2018a}.

Note that this fit procedure is different from that of Ref.~\cite{romerowolf2018a} as it accounts for detector effects. Nevertheless, the range of parameters scanned in our work is more narrow than in~\cite{romerowolf2018a}.
\section{Detailed results of the fit}\label{app:fitRes}

In Sec.~\ref{app:fitting} we have presented the method used in the fit. A summary of the main results were presented in Sec.~\ref{sec:fitResults}. In this appendix we provided additional information concerning the results.

\begin{figure}[ht!]
	\includegraphics[width=0.495\columnwidth]{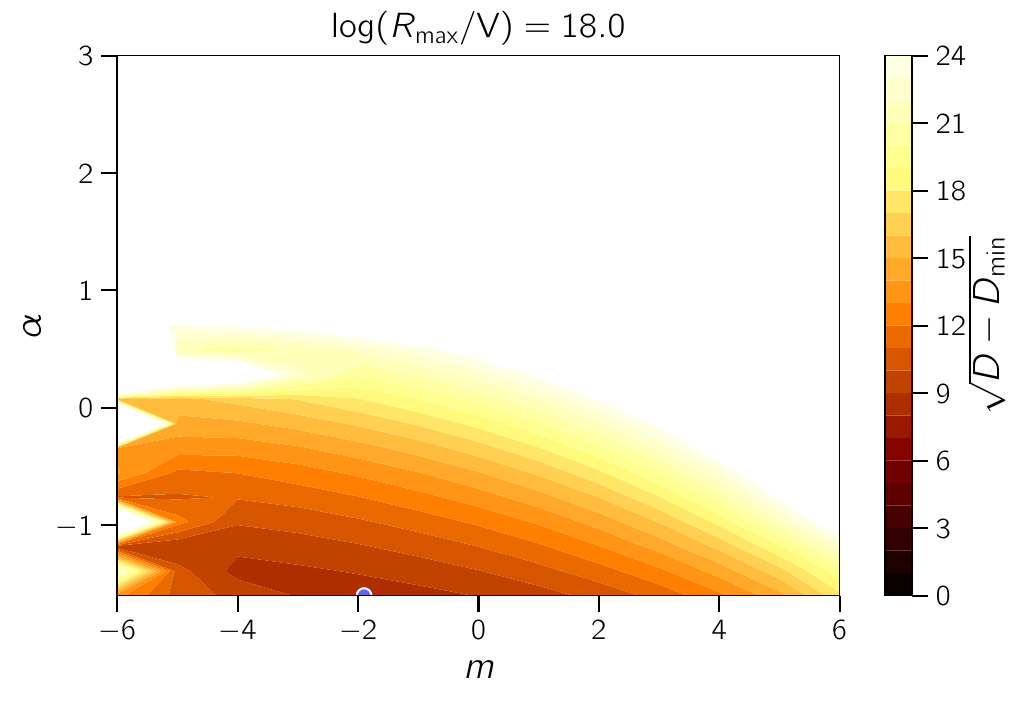}
	\includegraphics[width=0.495\columnwidth]{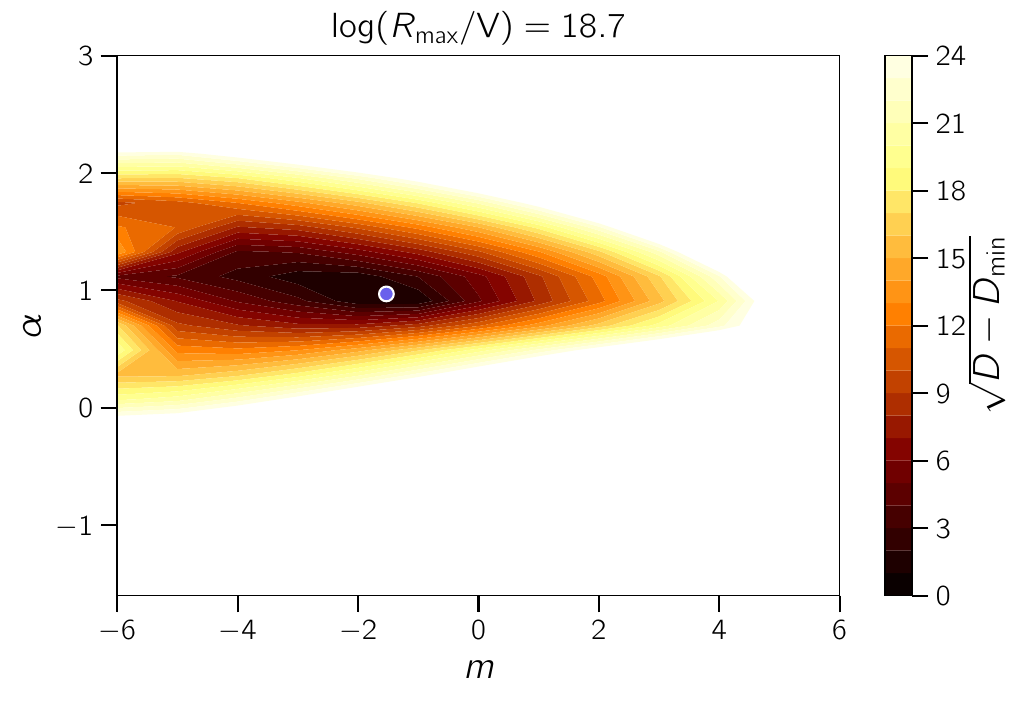}
	\includegraphics[width=0.495\columnwidth]{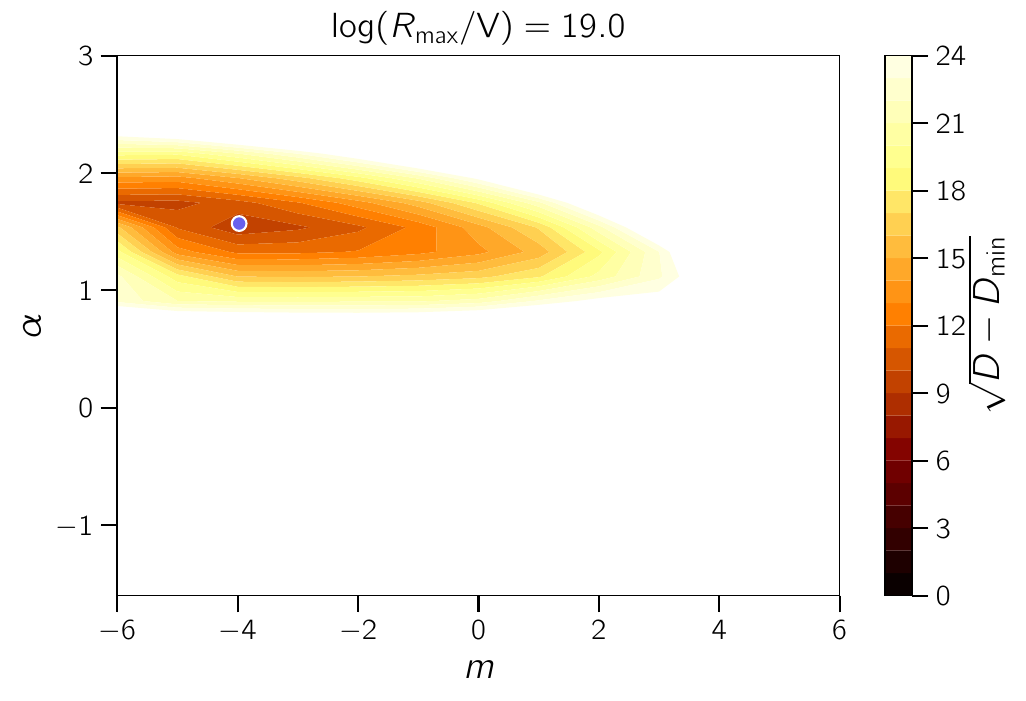}
	\includegraphics[width=0.495\columnwidth]{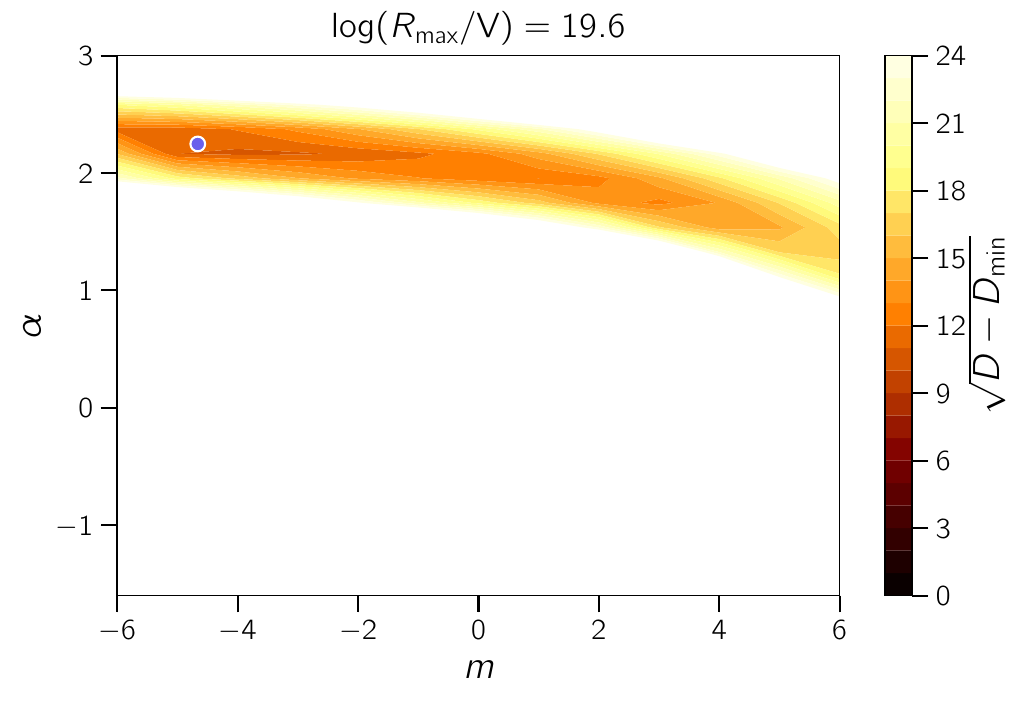}
	\caption{Parameter space of spectral index ($\alpha$) and source evolution ($m$) for maximal rigidities ($R_\text{max}$) $10^{18.0}$ (upper left), $10^{18.7}$ (upper right), $10^{19}$ (lower left), and $10^{19.6}$ V (lower right panel). The colour scale corresponds to $\sqrt{D - D_\text{min}}$. The circle indicates the best-fit parameters.}
	\label{fig:alpha_m}
\end{figure}

In Fig.~\ref{fig:alpha_m} we present the behaviour of the spectral index, $\alpha$, as a function of the source evolution parameter, $m$. A preference for lower values of $m$ can be clearly seen, further corroborating the results from Sec.~\ref{sec:fitResults}.

The results for each parameter, marginalised over the others, are listed in Table~\ref{tab:bestFitParameters}.

\begin{figure}[hbt!]
  \centering
  \includegraphics[width=0.50\columnwidth]{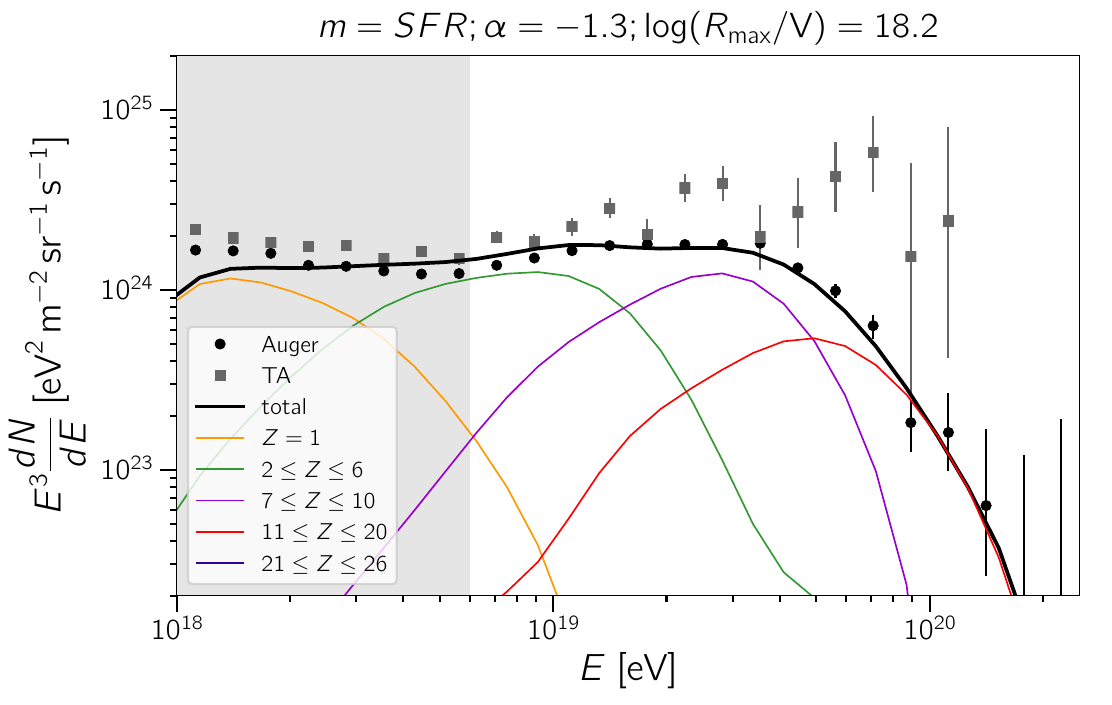} \\
  \includegraphics[width=0.495\columnwidth]{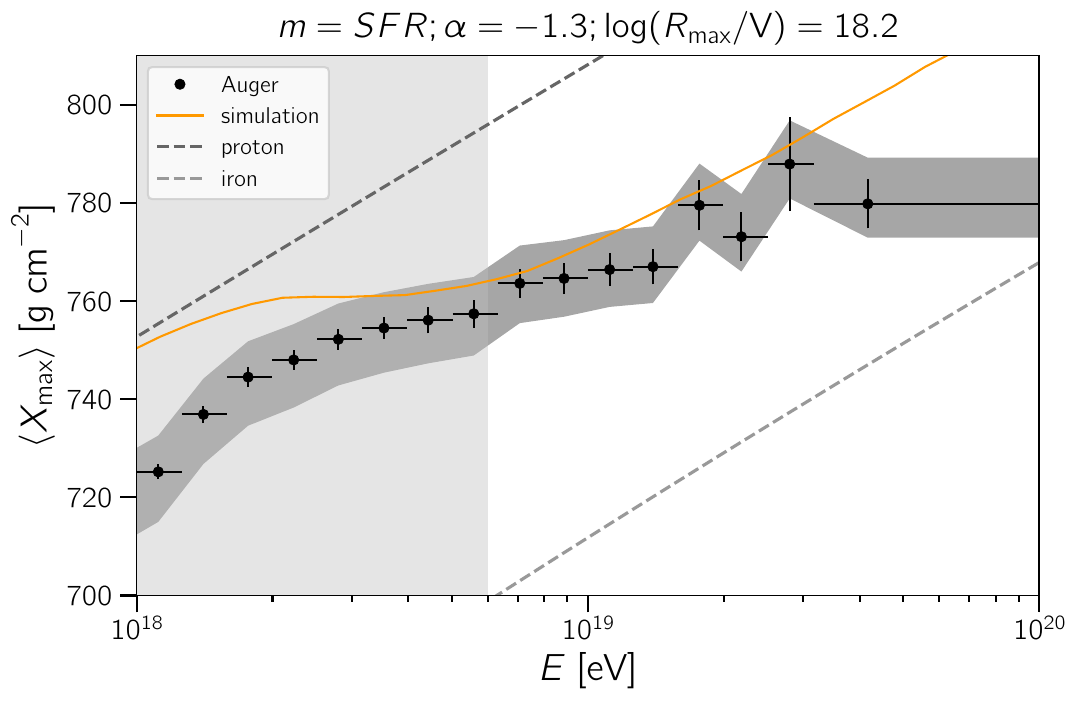}
  \includegraphics[width=0.495\columnwidth]{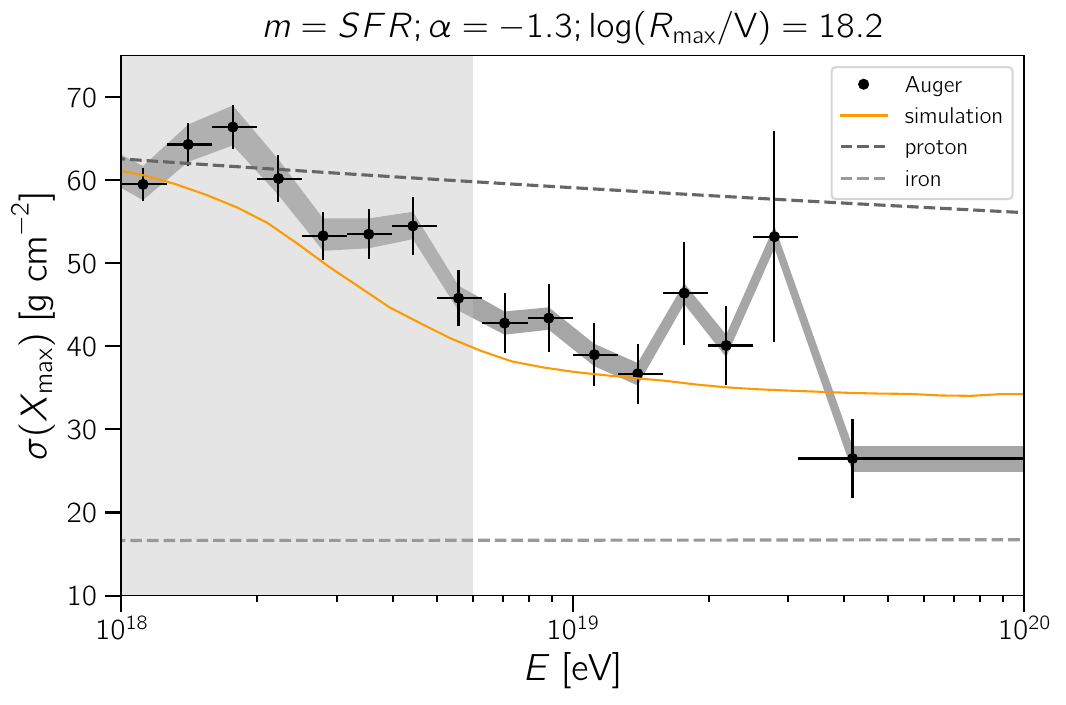}
  \caption{Spectrum (upper panel), $\langle X_\text{max} \rangle$ (lower left), and $\sigma(X_\text{max}$) (lower right panel) for the best fit to Auger data~\cite{auger2016b}, assuming that sources evolve as the star-formation rate. The spectrum measured by the Telescope Array~\cite{tinyakov2014a} is also shown for reference. The grey region at $E < 10^{18.7} \; \text{eV}$ refers to the energy range shown in the plots whose points were not used in fits. The dark grey regions around the composition-related observables correspond to the systematic uncertainties of Auger. The fractions of each injected element are: $f_\text{H}=0.1628$, $f_\text{He}=0.8046$, $f_\text{N}=0.0309$, $f_\text{Si}=0.0018$, and $f_\text{Fe}=0.0000$.}
  \label{fig:speccompSFR}
\end{figure}

\begin{figure}[hbt!]
  \centering
  \includegraphics[width=0.50\columnwidth]{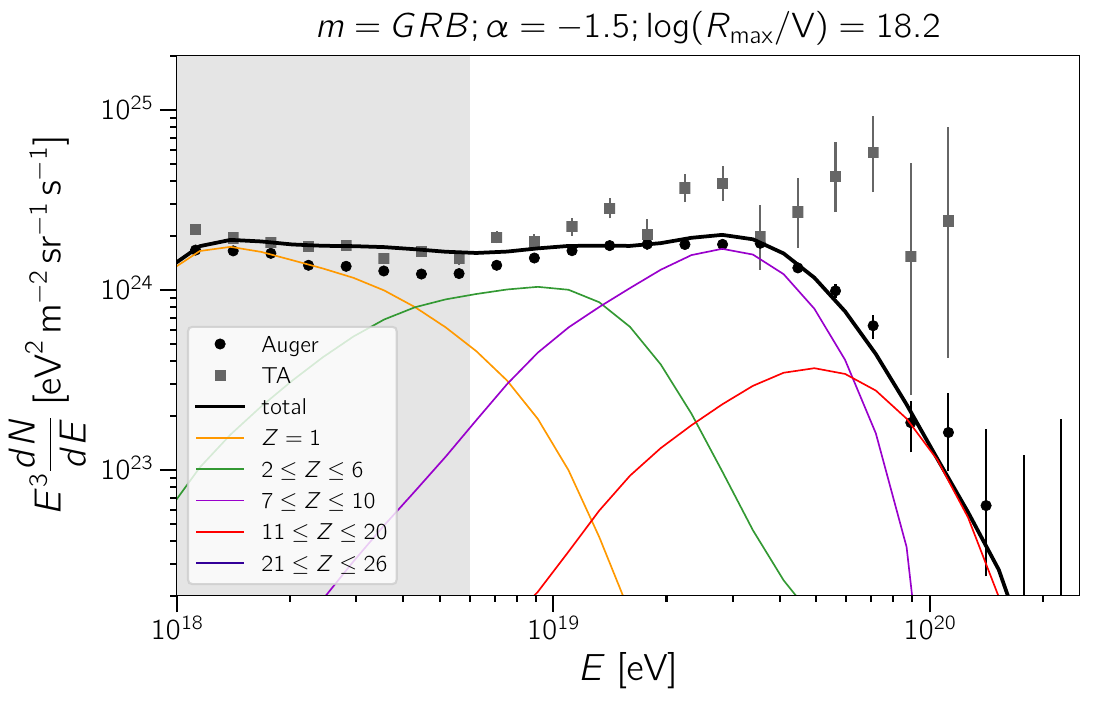} \\
  \includegraphics[width=0.495\columnwidth]{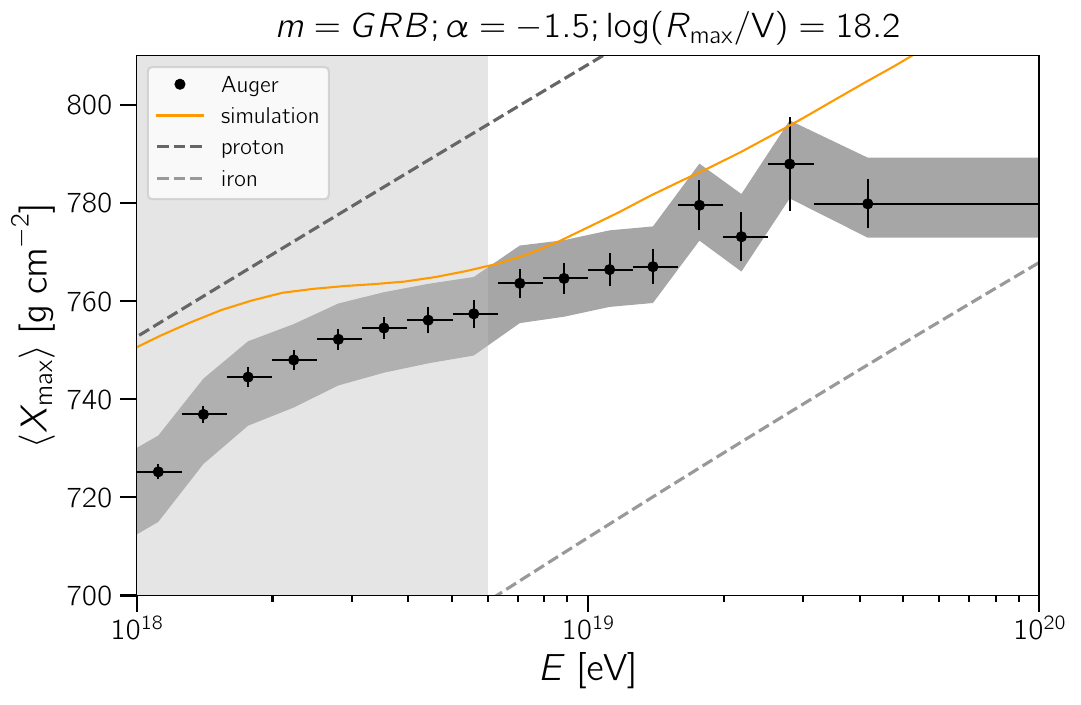}
  \includegraphics[width=0.495\columnwidth]{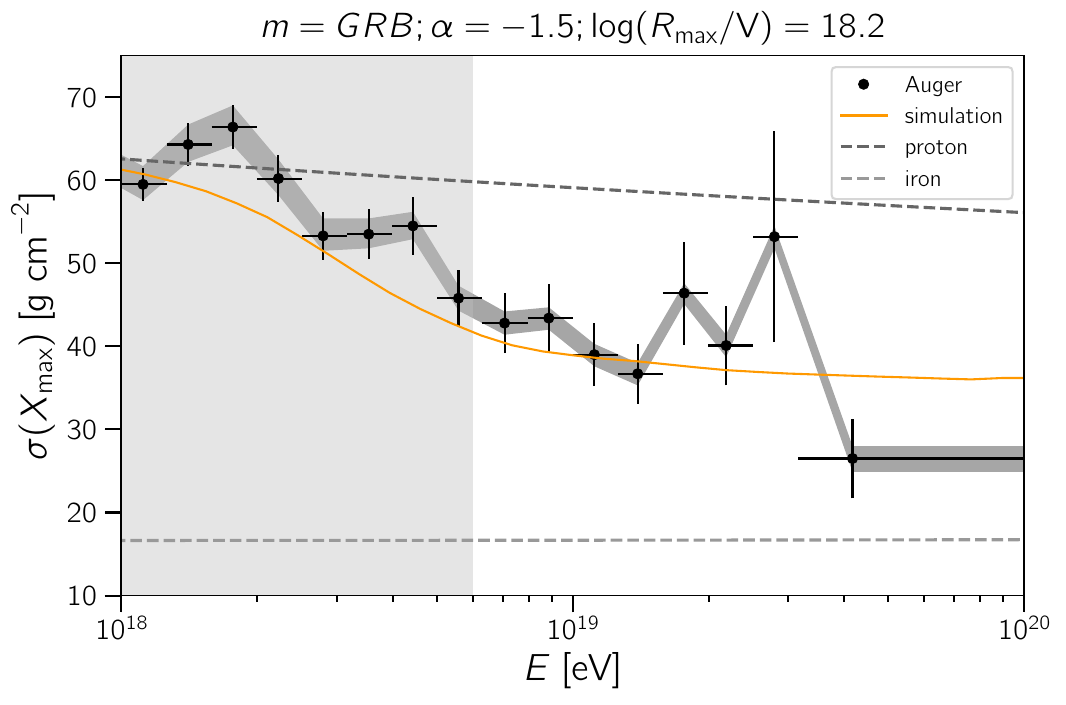}
  \caption{Sames as Fig.~\ref{fig:speccompSFR} assuming the GRB evolution. The fractions of each injected element are: $f_\text{H}=0.5876$, $f_\text{He}=0.3973$, $f_\text{N}=0.0147$, $f_\text{Si}=0.0004$, and $f_\text{Fe}=0.0$.}
  \label{fig:speccompGRB}
\end{figure}

\begin{figure}[hbt!]
  \centering
  \includegraphics[width=0.50\columnwidth]{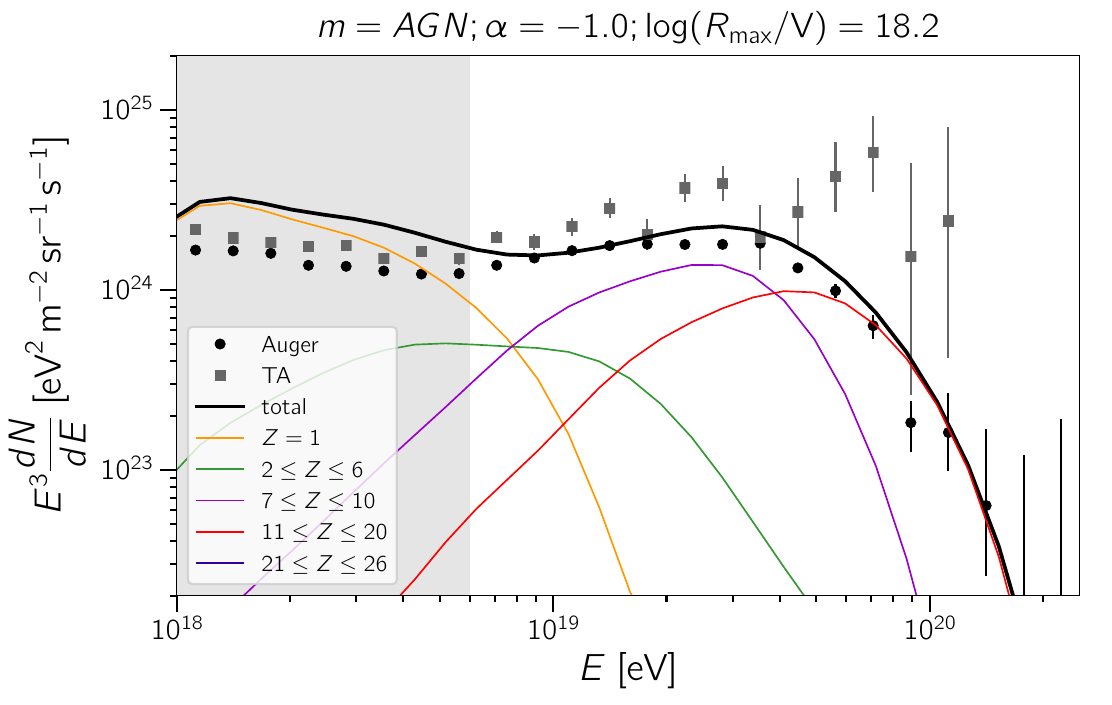} \\
  \includegraphics[width=0.495\columnwidth]{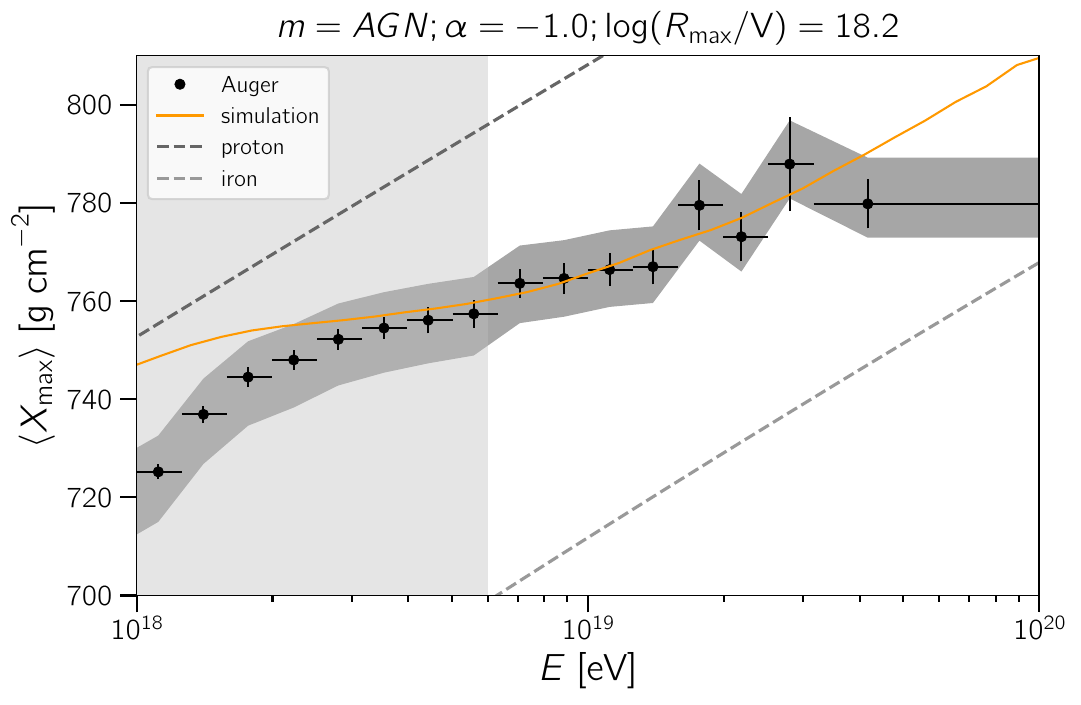}
  \includegraphics[width=0.495\columnwidth]{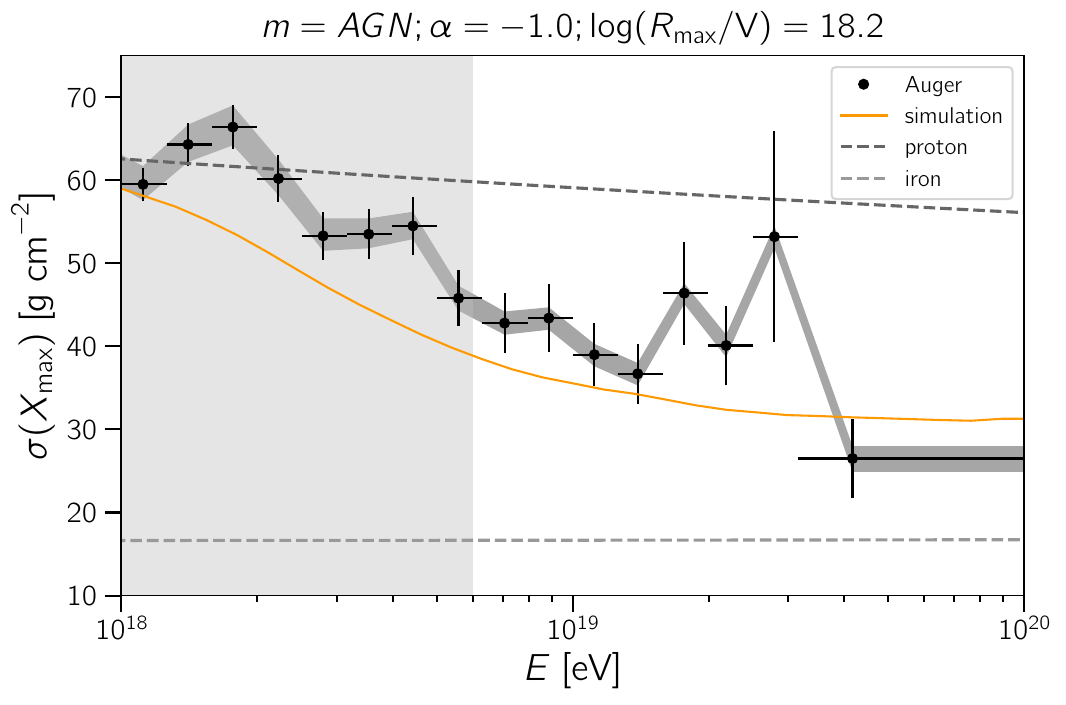}
  \caption{Same as Fig,~\ref{fig:speccompSFR} assuming the AGN evolution. The fractions of each injected element are: $f_\text{H}=0.8716$, $f_\text{He}=0.0778$, $f_\text{N}=0.0469$, $f_\text{Si}=0.0038$, and $f_\text{Fe}=0.0$.}
  \label{fig:speccompAGN}
\end{figure}

In Fig.~\ref{fig:speccomp} the best-fit results for spectrum and composition are shown for the scenarios with sources evolving as $(1 + z)^m$. The corresponding results for the SFR, GRB, and AGN cases are shown in Figs.~\ref{fig:speccompSFR},~\ref{fig:speccompGRB}, and~\ref{fig:speccompAGN}, respectively.

\section{Effects of the maximum redshift}\label{app:zmax}

The contribution of sources that are located farther than $z \sim 1$ to the UHECR spectrum at $E \gtrsim 10^{18.7} \; \text{eV}$ is virtually negligible. Therefore, our fit is not sensitive to this region. Nevertheless, the number of photons and neutrinos that are produced if one consider sources at $z \gtrsim 1$ highly depends on the maximal redshift set in the simulations, $z_\text{max}$. For this reason, we dedicate this appendix to discuss how this assumption would affect our results.

It should be noted that in this work we aim to provide \emph{conservative} predictions. Therefore, we restrict ourselves to $z<1$, since the contribution of sources at $z \gtrsim 1$ is virtually zero at $E > 10^{18.7} \; \text{eV}$. Any attempt to go beyond $z_\text{max} = 1$ would be an extrapolation and the cosmogenic fluxes predicted would not be completely reliable.  Alternative ways to model the emissivity evolution for $z > z_\text{max}$ would be to consider it flat for $z > z_\text{max}$, with a cutoff at a given redshift. 

There are no compelling physical arguments for using $z_\text{max} = 1$, as we have done, besides our inability to probe $z > 1$ with UHECRs above the ankle. If sources follow the star formation rate, for example, then they would peak at $z \simeq 2$; as a consequence, the contribution of object located at $1 \lesssim z \lesssim 2$ could be important. Similar arguments apply for most classes of astrophysical objects capable of accelerating cosmic rays to ultra-high energies. Exceptions are some models that predict UHECR production in tidal disruption events, for example. In this case, the interplay between the evolution of black holes and stars could yield an effective low UHECR emissivity at high redshifts.

We investigate the effects of the maximum redshift for two particular scenarios. As a benchmark, we adopt a spectrum with $\alpha = 1$ and $\log(R_\text{max} / \text{V}) = 20$. We consider two possible evolutions, $m=0$ and $m=3$, as the conclusions for $m < 0$ can be inferred from the comparison between these cases. We only study the scenarios of proton and helium injection, but the same arguments apply for any other nuclear species.

\begin{figure}[h!]
  \includegraphics[width=0.495\columnwidth]{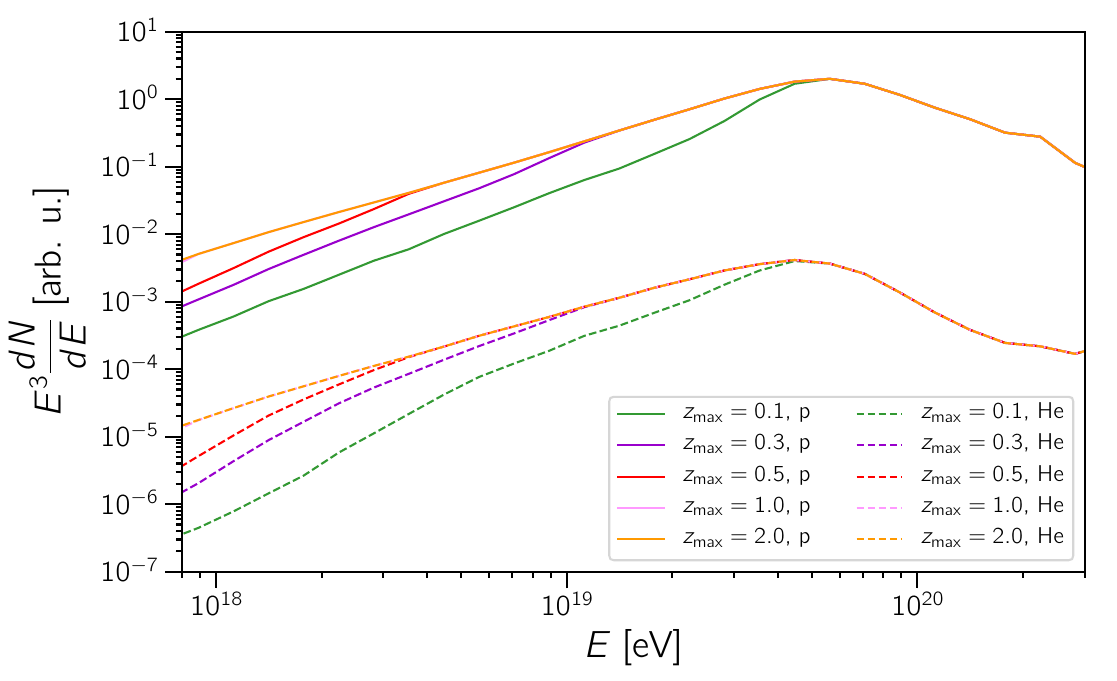}
  \includegraphics[width=0.495\columnwidth]{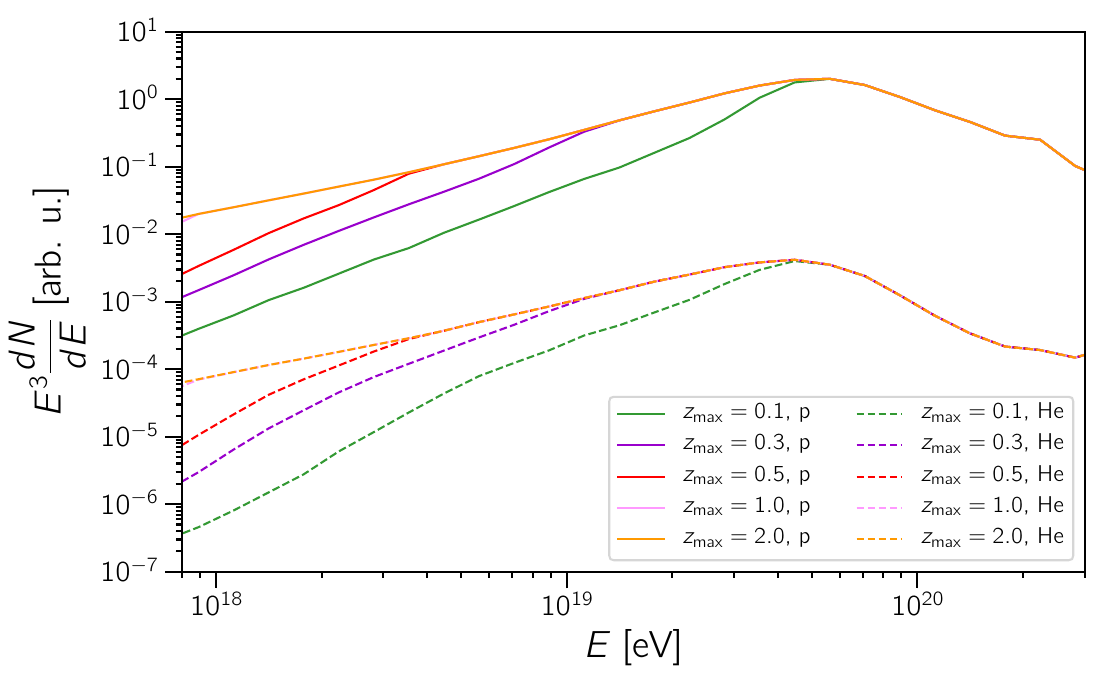}
  \includegraphics[width=0.495\columnwidth]{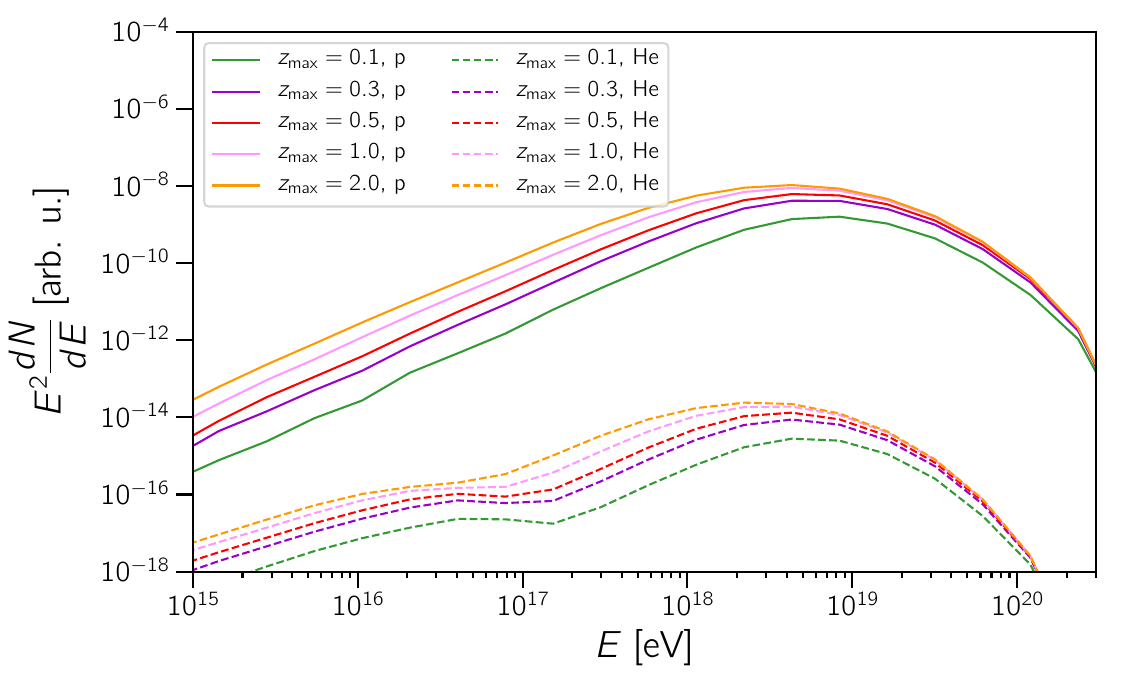}
  \includegraphics[width=0.495\columnwidth]{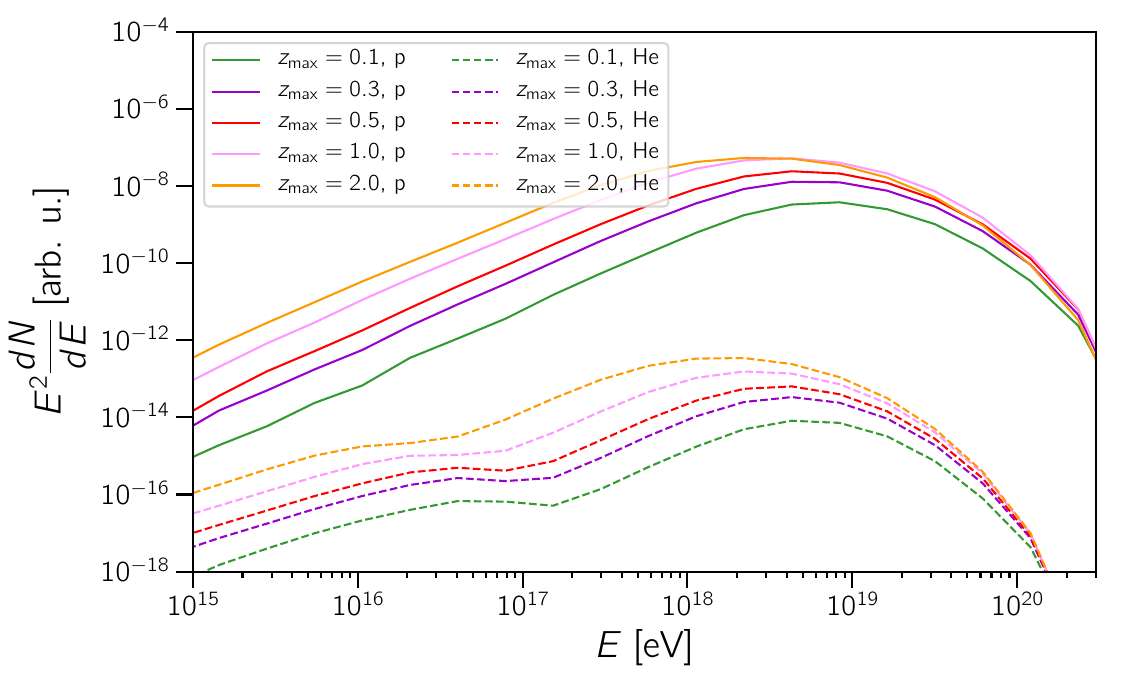}
  \includegraphics[width=0.495\columnwidth]{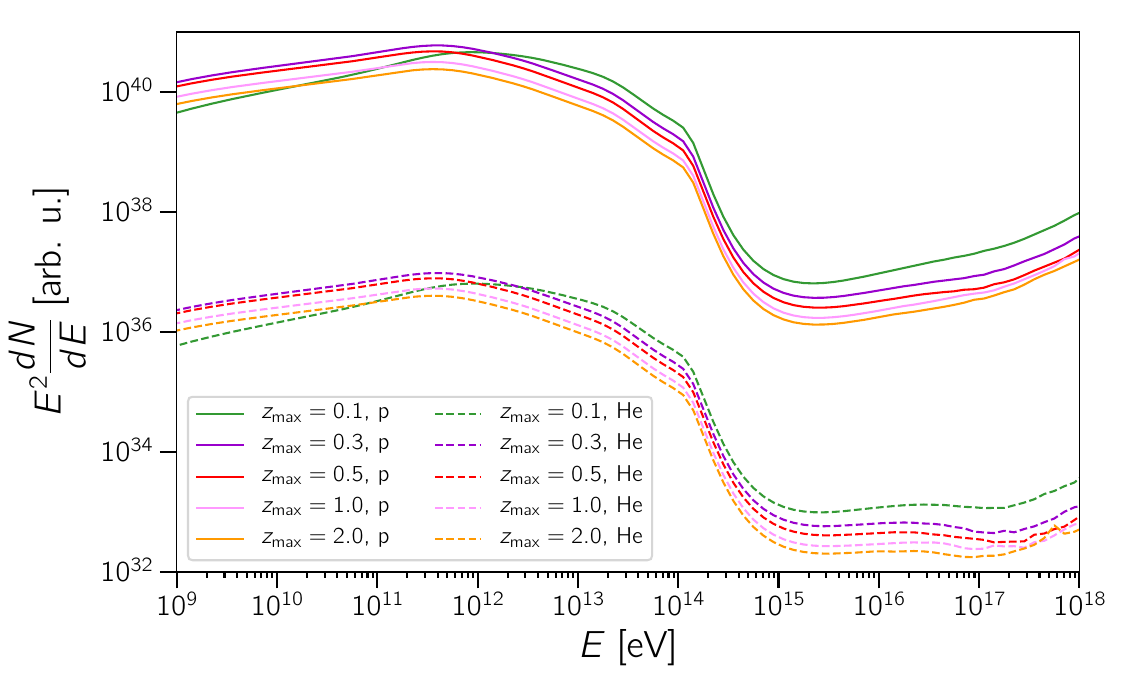}
  \includegraphics[width=0.495\columnwidth]{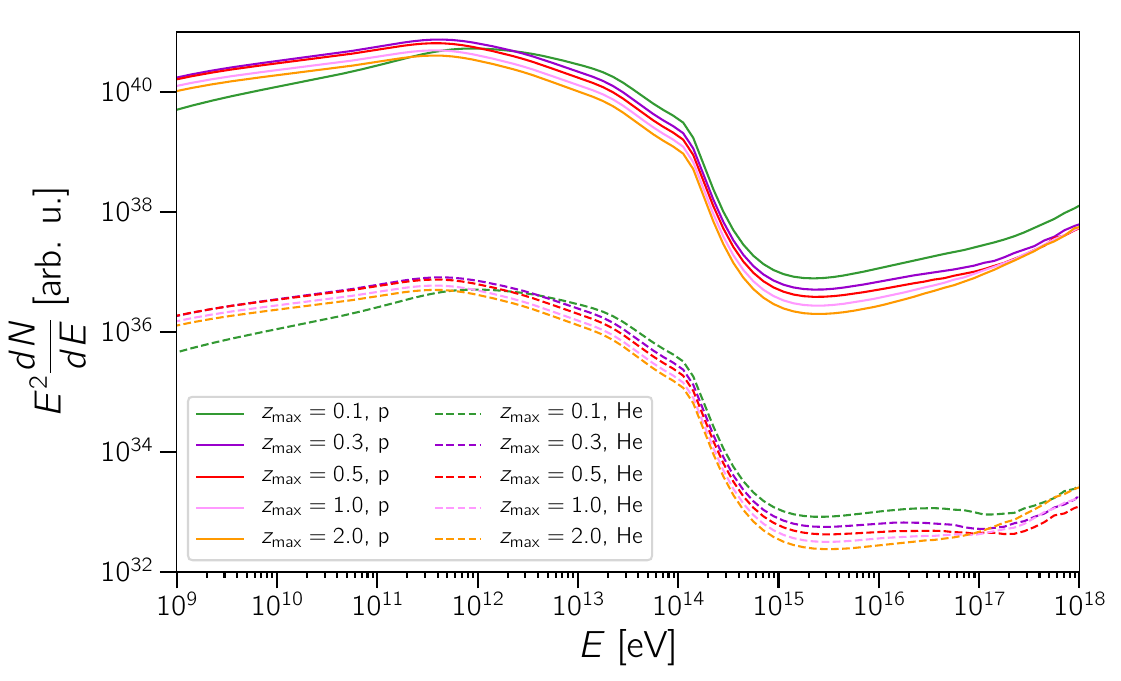}
  \caption{UHECR (upper row), neutrino (middle), and photon (lower row) for injected protons (solid lines) and helium nuclei (dashed lines). These scenarios are for $\alpha = 1$, $\log(R_\text{max} / \text{V}) = 20$. Evolution parameters $m=0$ are shown on the left column, and $m=3$ are shown on the right. The absolute normalisation of the fluxes for different nuclear species is arbitrarily chosen.}
  \label{fig:zmax}
\end{figure}

In Fig.~\ref{fig:zmax} we show the UHECR spectrum for proton and helium injection, along with the cosmogenic neutrino flux for different values of $z_\text{max}$. It is clear that if we increase $z_\text{max}$ from 1 to 2, the cosmogenic neutrino fluxes would increase by an order of magnitude for $m=0$ and even more for $m=3$.

Photons from distant sources are completely absorbed by the cosmological backgrounds (CMB, EBL, URB) at high energies. At low energies, most of the photons are from electromagnetic cascades induced by their higher-energy counterpart. Nevertheless, one can see that both for $m=0$ and $m=3$ the choice of $z_\text{max}$ does not have a major impact on the cosmogenic photon flux between GeV and PeV energies, being the difference between $z_\text{max}=1$ and $z_\text{max}=2$ only a factor of a few.

\begin{figure}[h!]
  \includegraphics[width=0.495\columnwidth]{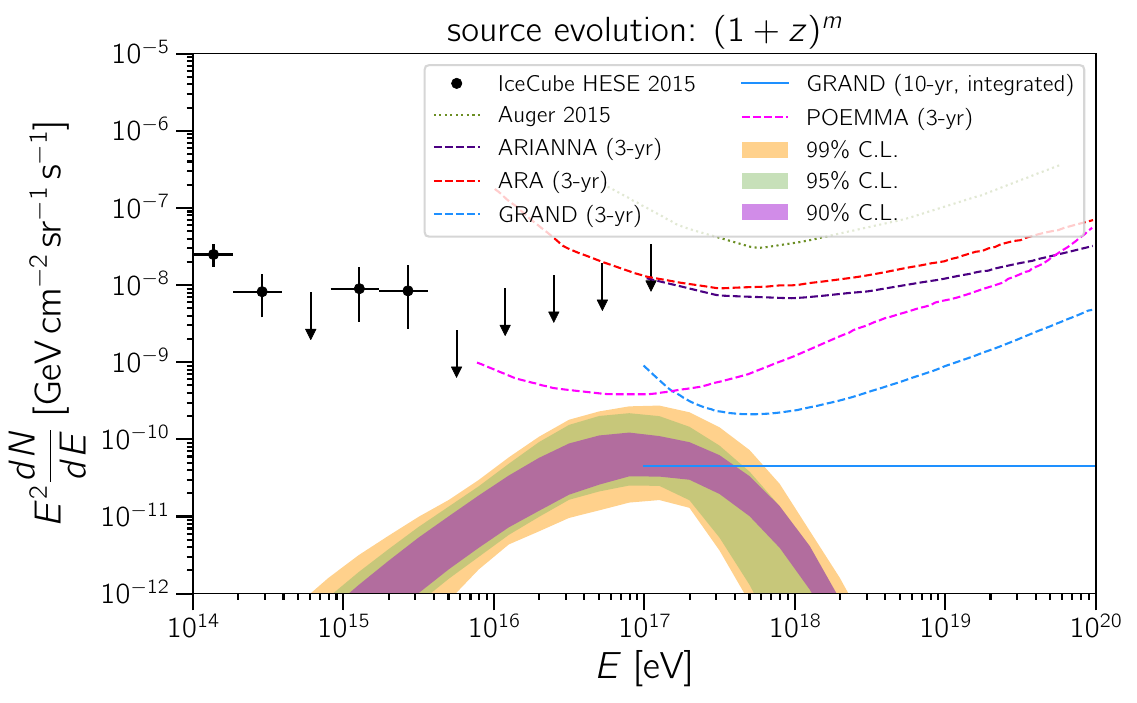}
  \includegraphics[width=0.495\columnwidth]{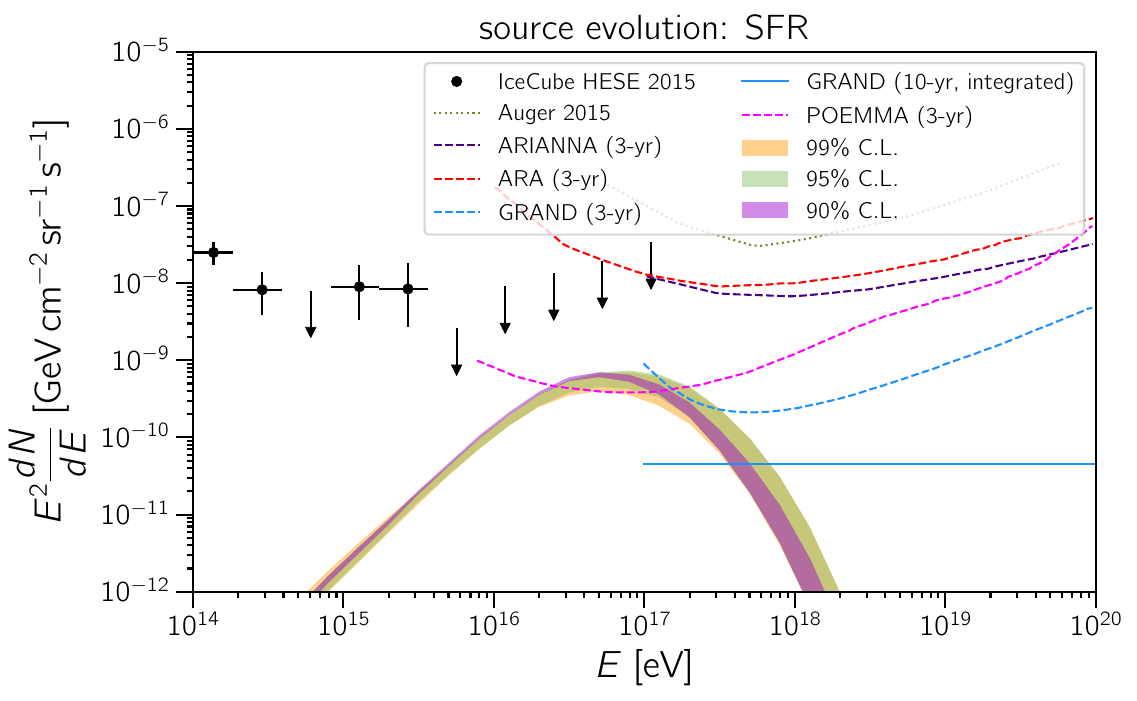}
  \includegraphics[width=0.495\columnwidth]{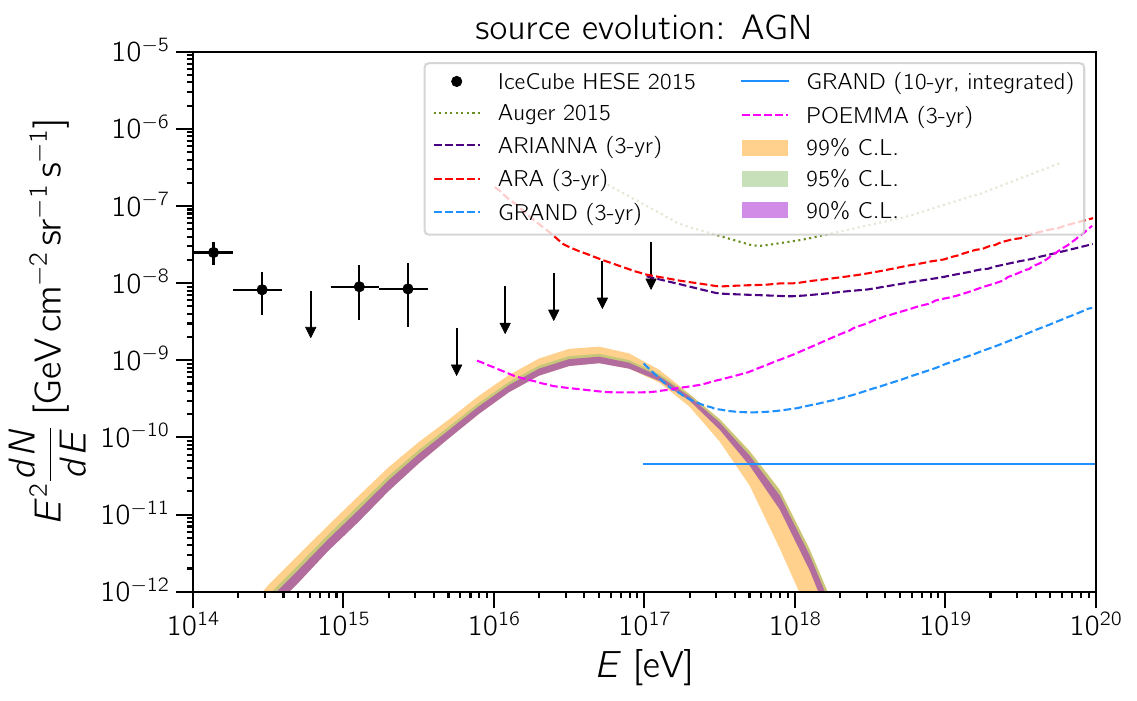}
  \includegraphics[width=0.495\columnwidth]{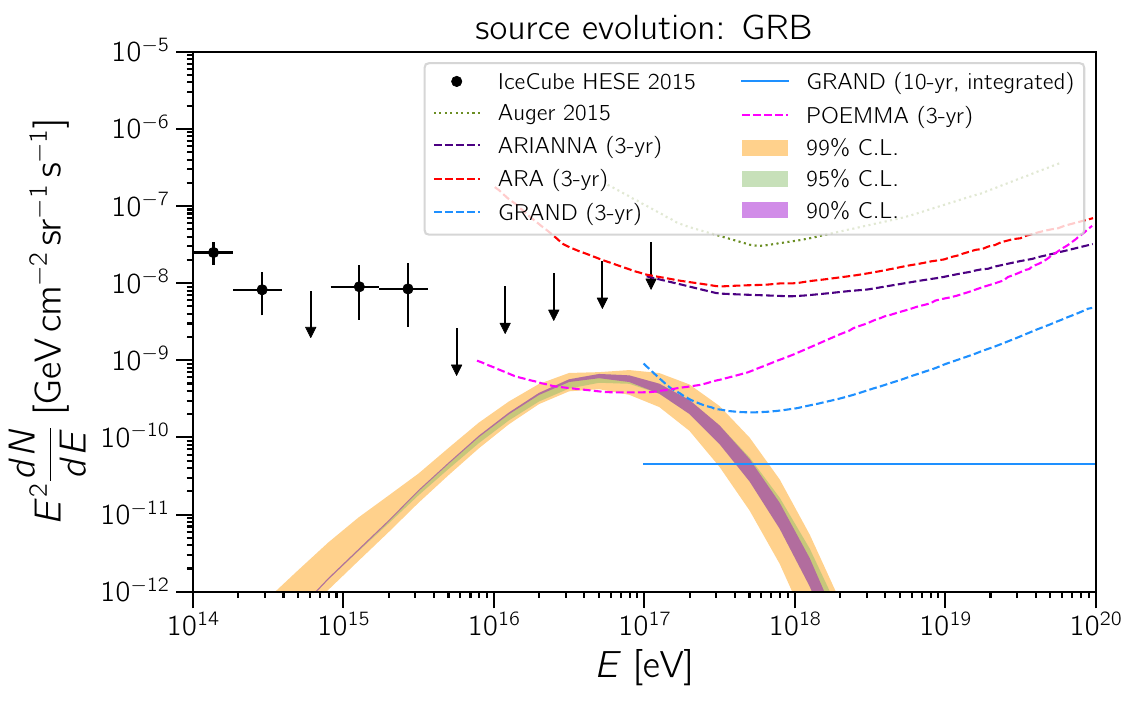}
  \caption{All-flavour ($\nu_e : \nu_\mu : \nu_\tau = 1:1:1$) cosmogenic neutrino fluxes for the best-fit scenarios with 90\%, 95\%, and 99\% confidence level. The sensitivity curves for Auger~\cite{auger2015a} (dotted lines) is shown, together with IceCube HESE events~\cite{icecube2015a} (black circles). The projected 3-year sensitivities for ARIANNA~\cite{arianna2015a}, ARA~\cite{ara2012a}, POEMMA~\cite{poemma2017a}, and GRAND~\cite{grand2017a,grand2017b,grand2018a} are also displayed as dashed lines. The upper left panel corresponds to a source evolution $(1 + z)^m$, the upper right to SFR, and the left and right lower rows are for AGN and GRB evolutions, respectively. These fluxes were computed assuming $z_\text{max}=5$ (see also Fig.~\ref{fig:cosmogenicNeutrinos}, for $z_\text{max}=1$).}
  \label{fig:cosmogenicNeutrinos2}
\end{figure}

In Fig.~\ref{fig:cosmogenicNeutrinos2} we present the cosmogenic fluxes assuming $z_\text{max}=5$. Note that the fluxes for the SFR, AGN, and GRB cases are significantly larger than the ones presented in Fig.~\ref{fig:cosmogenicNeutrinos} and are detectable by GRAND and POEMMA at a confidence level of 99\%. The scenario with evolution $(1 + z)^m$, however, still has a low flux of neutrinos due to the degeneracy between spectral index and source evolution parameter -- for small $\alpha$, the best fit is obtained for large $m$ and vice-versa.

\end{document}